\documentclass[letter,11pt]{article}
\usepackage{jheppub} 
                     
\pdfoutput=1

\usepackage[T1]{fontenc} 

\usepackage{slashed}
\usepackage{epsfig}
\usepackage{comment}
\usepackage{cancel}
\usepackage{bbm}
\usepackage{array}
\usepackage{bigints}
\usepackage{booktabs}
\usepackage{color}
\usepackage{dsfont}
\usepackage{float}
\usepackage{framed}
\usepackage{graphicx}
\usepackage{indentfirst}
\usepackage{mathrsfs}
\usepackage{multirow}
\usepackage{subdepth}
\usepackage{titlesec}
\usepackage[dotinlabels]{titletoc}
\usepackage{wrapfig}
\usepackage[all]{xy}
\usepackage[vcentermath]{youngtab}
\usepackage{relsize}
\usepackage{hyperref}
\numberwithin{equation}{section}

\usepackage[utf8]{inputenc}
\usepackage{slashed}
\usepackage{amsmath}
\usepackage{amsfonts}
\usepackage{amssymb}
\usepackage{mathtools}

\usepackage{cleveref}
\crefname{section}{§}{§§}
\Crefname{section}{§}{§§}

\def\e{{\epsilon}}

 \def\p{\partial}

\def\0{{(0)}}
\def\1{{(1)}}
\def\2{{(2)}}

\def\ci{{\mathscr I}}

\def\<{\langle }
\def\>{\rangle }

\def\l{{\lambda}}
\def\T{{\Theta}}
\def\ap{{\text{ap}}}
\def\G{{\Gamma}}
\def\cc{{\text{c.c.}}}
\def\o{{\omega}}
\def\k{{\kappa}}
\newcommand{\bea}{\begin{eqnarray}}
\newcommand{\eea}{\end{eqnarray}}
\newcommand{\be}{\begin{equation}}
\newcommand{\ee}{\end{equation}}
\newcommand{\ba}{\begin{align}}
\newcommand{\ea}{\end{align}}

\renewcommand{\O}{\Omega}

\newcommand{\bigO}{\mathcal{O}}

\newcommand{\w}[1]{\mbox{$\W_\infty[#1]$}}

   \makeatletter
  \let\over=\@@over \let\overwithdelims=\@@overwithdelims
  \let\atop=\@@atop \let\atopwithdelims=\@@atopwithdelims
  \let\above=\@@above \let\abovewithdelims=\@@abovewithdelims
\renewcommand\section{\@startsection {section}{1}{\z@}%
                                   {-3.5ex \@plus -1ex \@minus -.2ex}
                                   {2.3ex \@plus.2ex}%
                                   {\normalfont\large\bfseries}}

\renewcommand\subsection{\@startsection{subsection}{2}{\z@}%
                                     {-3.25ex\@plus -1ex \@minus -.2ex}%
                                     {1.5ex \@plus .2ex}%
                                     {\normalfont\bfseries}}

\newcommand{\beq}{\begin{equation}}
\newcommand{\eeq}{\end{equation}}
\newcommand{\beqa}{\begin{eqnarray}}
\newcommand{\eeqa}{\end{eqnarray}}
\newcommand{\beqar}{\begin{eqnarray*}}

\newcommand{\ve}{{\varepsilon}}

\def\[{\big[}
\def\]{\big]}

\def\e{{\epsilon}}
\def\ve{{\varepsilon}}

\def\g{{\gamma}}
\def\o{{\omega}}
\def\a{{\alpha}}

\def\t{{\theta}}
\def\O{\Omega}

\def\w{\omega}

\def\be{{\bar \epsilon}}

\def\CA{{\mathcal A}}

\def\CI{{\mathcal I}}
\def\ci{{\mathscr I}}

\def\CO{{\mathcal O}}

\def\CS{{\mathcal S}}

\def\SO{{\mathscr O}}

\newcommand{\bra}[1]{\langle\,   #1\,    |}
\newcommand{\ket}[1]{ |\,   #1 \,  \rangle}
\newcommand{\braket}[2]{\langle\,   #1 \, | \, #2 \, \rangle}

\def\mrr{{\mathbb R}}
\def\mzz{{\mathbb Z}}

\def\pmm{{(\pm)}}

\def\+{{(+)}}
\def\-{{(-)}}
\def\0{{(0)}}
\def\1{{(1)}}
\def\2{{(2)}}
\def\3{{(3)}}

\title{\boldmath Asymptotic Symmetries and Weinberg's Soft Photon Theorem in Mink$_{d+2}$}

\author[]{Temple He$^1$}
\author[]{and Prahar Mitra$^2$}

\affiliation[]{$^1$Center for Quantum Mathematics and Physics (QMAP), University of California, Davis, CA 95616, USA}\affiliation[]{$^2$School of Natural Sciences, Institute for Advanced Study, Princeton, NJ 08540, USA}

\emailAdd{tmhe@ucdavis.edu}
\emailAdd{prahar21@ias.edu}

\abstract{We show that Weinberg's leading soft photon theorem in massless abelian gauge theories implies the existence of an infinite-dimensional large gauge symmetry which acts non-trivially on the null boundaries $\ci^\pm$ of $(d+2)$-dimensional Minkowski spacetime. These symmetries are parameterized by an arbitrary function $\ve(x)$ of the $d$-dimensional celestial sphere living at $\ci^\pm$. This extends the previously established equivalence between Weinberg's leading soft theorem and asymptotic symmetries from four and higher even dimensions to \emph{all} higher dimensions.}

\begin{document} 
\maketitle
\flushbottom


\section{Introduction}

Soft theorems are equations that describe how scattering amplitudes in Minkowski spacetime involving one or many low energy, or soft, massless particles factorizes into a term involving the soft gauge particles (known as the soft factor) and one involving the amplitude without the soft particles, i.e.
\begin{equation}
\begin{split}
	\CA_n ~~ \xrightarrow{m~\text{soft particles}} ~~ S_m  \CA_{n-m} ,
\end{split}
\end{equation}
where $S_m$ is the soft factor involving $m$ soft particles. Such theorems were established over half a century ago in the works of Bloch, Nordseick, Low, Yennie, Frautschi, Suura \cite{Bloch:1937pw,Nordsieck:1937zz,Low:1954kd,GellMann:1954kc,Low:1958sn,Yennie:1961ad} and in a more modern language by Weinberg \cite{Weinberg:1965nx,Weinberg:1995mt}. A particularly interesting feature of all the soft theorems is that the soft factor is \emph{universal}, i.e. it does not depend on the details of the theory under consideration. Rather, it depends only on the quantum numbers (linear momenta, angular momenta and charges) of the external particles involved in the scattering amplitude. Such universality naturally suggests that perhaps the soft theorems arise from some underlying symmetry, and indeed, it was relatively recently discovered that these theorems are precisely the Ward identities associated to the asymptotic symmetries of scattering amplitudes (see \cite{Strominger:2017zoo} for a review and a comprehensive list of references).

Largely in part due to this discovery, in the last several years, there has been a resurgence of activity in the study of soft theorems, and many interesting directions have been explored. The equivalence was originally conceived for four-dimensional theories where it has been established to hold for gauge theories \cite{He:2014cra,Mohd:2014oja,He:2015zea,Campiglia:2015qka,Kapec:2015ena,Strominger:2015bla,Gabai:2016kuf}, gravity \cite{He:2014laa,Campiglia:2015yka,Campiglia:2015kxa}, as well as their supersymmetric counterparts \cite{Dumitrescu:2015fej}. The equivalence has proved useful in the exploration of not just the original soft theorems, but also new subleading and sub-subleading soft theorems \cite{Cachazo:2014fwa,Schwab:2014fia,Kalousios:2014uva,Luo:2014wea,DiVecchia:2016amo,Sen:2017nim,Laddha:2017vfh,Hirai:2018ijc}. Some of these were discovered by studying known asymptotic symmetries \cite{Barnich:2011ct,Kapec:2014opa,Campiglia:2014yka} while others led to the discovery of new asymptotic symmetries \cite{Lysov:2014csa,Campiglia:2016jdj,Campiglia:2016hvg,Conde:2016csj,Campiglia:2016efb,Conde:2016rom,Laddha:2017ygw}. Finally, this equivalence has also been extended to theories living in higher even dimensions \cite{Schwab:2014xua,Kapec:2014zla,Kapec:2015vwa}, which is certainly relevant in the context of string theory.\footnote{More generally, soft theorems and asymptotic symmetries are also equivalent to memory effects which have been explored in \cite{Thorne:1992sdb,Strominger:2014pwa,Pasterski:2015zua,Bieri:2015yia,Susskind:2015hpa,Hotta:2016qtv,Hollands:2016oma,Pate:2017vwa,Pate:2017fgt,Ball:2018prg}, but this will be beyond the scope of this paper (see \cite{Strominger:2017zoo} for a review).}

The plethora of successes in these endeavors has given confidence that perhaps every soft theorem in Minkowski spacetime can be interpreted as the Ward identity for some asymptotic symmetry. However, there is a noticeable class of theories in which the bridge between soft theorems and asymptotic symmetries is missing -- these are theories in odd spacetime dimensions. Soft theorems are true and have the same universal form in both even and odd dimensions.\footnote{The standard derivation of these theorems in perturbation theory relies largely on the fact that the free field propagator has the form $p^{-2}$, which is true in any dimension.} Yet, to our knowledge, there has been no successful attempt in showing the equivalence of soft theorems and asymptotic symmetries in odd dimensional theories. The fundamental issue lies in the qualitatively different properties of massless waves in even and odd dimensions. For instance, the massless Green's function is supported entirely on the light cone in even dimensions, whereas in odd dimensions it is supported everywhere inside the lightcone \cite{Balazs1955,Soodak1993}. This implies that for a complete understanding of waves in odd dimensions, one needs to study timelike infinity $i^\pm$ as well as null infinity $\ci^\pm$. Another symptom of the same feature is that the near $\ci^\pm$ expansions of massless fields is analytic in even dimensions (i.e. admits a Taylor series expansion in $r^{-1}$ where $r$ is the radius of the transverse sphere) and non-analytic in odd dimensions. These issues make the discussion of asymptotic symmetries particularly difficult in odd dimensions. Despite this, it would be very strange if no such relationship exists, implying that soft theorems in odd dimensions have a wholly different origin than those in even dimensions.

In this paper, we resolve the issues discussed in the previous paragraph and show that, at least in the simple example of a $U(1)$ gauge theory with massless charged matter, the leading soft photon theorem is the Ward identity associated to large gauge symmetry in \emph{all} dimensions, thereby unifying and generalizing the previous discussions from four and higher even dimensions to all dimensions. We organize the paper so that in \S\ref{sec:asymptotics}, we study $U(1)$ gauge theories in $d+2$ dimensions and discuss asymptotic fall-offs of both free and interacting gauge fields. In \S\ref{sec:softphotonthm}, we impose a matching condition across spatial infinity $i^0$ and demonstrate that the resulting Ward identity implies the leading soft photon theorem. Finally, in \S\ref{sec:largegaugesym}, we show that the matching condition implies conservation of a charge which generates large gauge transformations on the fields at $\ci^\pm$.

\section{$U(1)$ Gauge Theories in $d+2$ Dimensions}\label{sec:asymptotics}

In this section, we introduce the conventions used in this paper and discuss the necessary background information regarding the equations of motion and the asymptotics of the $U(1)$ gauge field.

\subsection{Classical Equations}

We work in flat null coordinates $x^\mu = (u,r,x^a)$, $a=1,\ldots,d$ which are related to Cartesian coordinates $X^A$ by\footnote{These coordinates were utilized in the case of four dimensions in \cite{Dumitrescu:2015fej} and in \cite{Kapec:2017gsg} in higher dimensions.}
\begin{align}
 	X^A = \frac{r}{2}\left( 1+x^2 + \frac{u}{r},2x^a,1-x^2 - \frac{u}{r}\right), \qquad x^2 = \delta_{ab}x^ax^b.
\end{align}
Lowercase Latin indices are raised and lowered by the Kronecker delta $\delta^{ab}$ and $\delta_{ab}$, respectively. Using these coordinates, the metric of Minkowski spacetime becomes
\begin{equation}
\begin{split}
d s^2 &= \eta_{AB}dX^AdX^B =  - d u dr + r^2\delta_{ab} d x^a d x^b ,
\end{split}
\end{equation}
where $\eta_{AB} = (-1,1,\ldots,1)$ is the standard Minkowski metric in Cartesian coordinates. The null boundaries $\ci^\pm$ are located at $r \to \pm \infty$ while keeping $(u,x^a)$ fixed. On this hypersurface, $x^a$ is the stereographic coordinate on the celestial sphere. The point labeled by $x^a$ on $\ci^+$ is antipodal to the point with the same label on $\ci^-$. The boundaries of $\ci^+$ ($\ci^-$) are at $u \to \pm \infty$ and are denoted by $\ci^+_\pm$ ($\ci^-_\pm$). In Appendix \ref{sec:coordinates}, we provide more details about this coordinate system.

A $U(1)$ gauge theory is described by a field strength $F_{\mu\nu}$ and matter fields $\Psi_i $. The matter fields couple to the field strength via a conserved $U(1)$ current $J_\mu$. Dynamics of the field strength is described by Maxwell's equations and the Bianchi identity, given respectively by
\begin{align}
\label{maineq1} e^2 J_\nu &= \nabla^\mu F_{\mu\nu}    \\
\label{maineq2} 0 &= \p_{\a} F_{\mu\nu}  + \p_{\nu} F_{\a\mu}  + \p_{\mu} F_{\nu\a}  . 
\end{align}
In flat null coordinates, \eqref{maineq1} takes the form
\begin{align}\label{MaxwellEq}
\begin{split}
	e^2 J_u &= 2\p_u F_{ur} - \frac{1}{r^2}\p^a F_{ua} \\
	e^2 J_r &= -2\p_r F_{u r} - \frac{1}{r^2}\p^aF_{ra} - \frac{2d}{r} F_{ur}  \\
	e^2 J_a &= -2\p_r F_{u a} - 2\p_u F_{r a}  - \frac{2(d-2)}{r}F_{ua} - \frac{1}{r^2}\p^b F_{ab} .
\end{split}
\end{align}
Furthermore, the Bianchi identity \eqref{maineq2} is solved by
\begin{equation}
\begin{split}
F_{\mu\nu}  = \p_\mu A_\nu  - \p_\nu A_\mu  , \qquad A_\mu  \sim A_\mu + \p_\mu \ve   , \qquad \ve \sim \ve + 2 \pi . 
\end{split}
\end{equation}
Note that the theory is invariant under gauge transformations 
\begin{equation}
\begin{split}
A_\mu  ~\to~ A_\mu  + \p_\mu \ve  , \qquad \Psi_i  ~\to~ e^{ i Q_i \ve  } \Psi_i  ,
\end{split}
\end{equation}
where $Q_i \in \mzz$ is the $U(1)$ charge of the field $\Psi_i $.  Gauge transformations that vanish at infinity correspond to redundant descriptions of the same physical state and can be eliminated by a choice of gauge.

\subsection{Asymptotics}

In this section, we study the structure of the field strength near $\ci^\pm$. The general solution to \eqref{maineq1} and \eqref{maineq2} has the form
\begin{equation}
\begin{split}
F_{\mu\nu}  = F_{\mu\nu}^{(R)}  + F_{\mu\nu}^{(C)}  . 
\end{split}
\end{equation}
The radiative field $F_{\mu\nu}^{(R)} $ satisfies the homogeneous Maxwell's equations ($J_\nu=0$) and $F_{\mu\nu}^{(C)} $ satisfies \eqref{maineq1}. The separation of the field strength into a radiative part and a Coulombic part is fixed by either choosing a Green's function or by imposing boundary conditions on the fields. We will be interested in incoming $\-$ and outgoing $\+$ solutions to Maxwell's equations, whose radiative and Coulombic parts are denoted respectively by $F_{\mu\nu}^{(R\pm)}$ and $F_{\mu\nu}^{(C\pm)}$. These fields have the following large $r$ fall-offs near $\ci^\pm$:\footnote{As we shall see in \eqref{Furlargerodd1}, in odd dimensions, the radiative field has an expansion in both integer and half-integer powers of $|r|$, as indicated in the first column of \eqref{Ffalloff}. In even dimensions, we obtain later in \eqref{Furlargereven1} that the radiative field has an expansion in integer powers of $|r|$, though non-analytic terms of the form $\log r$ are also present (not shown in \eqref{Ffalloff}).}
\begin{equation}
\begin{split}\label{Ffalloff}
F_{ur}^{(R\pm)} &= O \left( |r|^{-\frac{d}{2}-1} \right)  + O \left( |r|^{-d}  \right)  , \qquad\quad F_{ur}^{(C\pm)} = O \left( |r|^{-d}  \right)  \\
F_{ra}^{(R\pm)} &= O \left( |r|^{-\frac{d}{2}} \right)  + O \left( |r|^{-d} \right)  , \qquad\qquad F_{ra}^{(C\pm)} = O \left( |r|^{-d} \right)  \\
F_{ua}^{(R\pm)} &= O \left( |r|^{-\frac{d}{2}+1} \right)  + O \left( |r|^{-d+1} \right)  , \qquad F_{ua}^{(C\pm)} = O \left( |r|^{-d+1} \right)  \\
F_{ab}^{(R\pm)} &= O \left( |r|^{-\frac{d}{2}+1} \right)  + O \left( |r|^{-d+1} \right)  , \qquad F_{ab}^{(C\pm)} = O \left( |r|^{-d+1} \right)  .
\end{split}
\end{equation}
The leading fall-offs for the radiative part are determined by noting that the electromagnetic stress tensor components $T_{uu}$ and $T_{ua}$ must fall off as $|r|^{-d}$ in order to have finite energy-momentum and angular-momentum flux through $\ci^\pm$. Then, using the fact that $T_{uu} \sim r^{-2} F_{ua} F_u{}^a$ and $T_{ua} \sim r^{-2} F_{ab} F_{u}{}^b$, we can determine the fall-offs for $F_{ua}$ and $F_{ab}$. The remaining ones are then determined by Lorentz invariance. Similarly, the fall-offs for the Coulombic part are determined by Gauss's law, which also determines the fall-offs for the matter current
\begin{equation}\label{currfalloff}
\begin{split}
J_u = O\big( |r|^{-d} \big)  , \qquad J_a = O \big( |r|^{-d} \big) , \qquad J_r = O \big( |r|^{-d-2} \big) .
\end{split}
\end{equation}
We would now like to determine the precise large $r$ expansion for both the radiative and the Coulombic fields.

\subsubsection{Radiative Field}\label{sec:radasymptotics}

The radiative gauge field can be decomposed into creation and annihilation operators $\bigO^{(\pm)\dag}_a$ and $\bigO^{(\pm)}_a$ as follows:
\begin{equation}
\begin{split} \label{gaugefieldmodeexp}
F^{(R\pm)}_{AB} (X) = e \int \frac{d^{d+1} q}{ ( 2\pi )^{d+1} } \frac{1}{2q^0} \left[ \ve^a_{AB} ({\vec q}\,) \CO^\pmm_a(\vec{q}\,) e^{ i q \cdot X }  +  \ve^a_{AB} ({\vec q}\,)^* \CO^{\pmm\dagger}_a(\vec{q}\,) e^{ - i q \cdot X }  \right] ,
\end{split}
\end{equation}
where $q^0 = | \vec{q}\,|$ and $\ve_{AB}^a(\vec{q}\,)$ are the $d$ polarization tensors labeled by $a$ and defined via 
\begin{equation}
\begin{split}\label{polvec}
\ve^a_{AB}(\vec{q}\,) = i \left[ q_A \ve^a_B({\vec q}\,)  - q_B \ve^a_A({\vec q}\,)  \right] , \qquad q^A \ve_A^a ({\vec q}\,) = 0 , \qquad \eta^{AB}\ve^{a}_A({\vec q}\,) \ve_B^b({\vec q}\,)^* =  \delta^{ab} .
\end{split}
\end{equation}
In a quantum theory, the creation and annihilation modes satisfy the canonical commutation relation
\begin{equation}
\begin{split}
\left[ \CO^\pmm_a (\vec{q}\,) \,,\, \CO^{\pmm\dagger}_b (\vec{q}\,') \right] = \big( 2 q^0  \big) \delta_{ab} (2\pi)^{d+1} \delta^{(d+1)}  ( \vec{q} - \vec{q}\,'  ) .
\end{split}
\end{equation}
The equations in \eqref{polvec} remain invariant under $\ve^a_A(\vec{q}\,) \to \ve^a_A(\vec{q}\,) + f^a(\vec{q}\,) q_A$ for any arbitrary functions $f^a$. We use this to parameterize the on-shell photon momentum and polarization as\footnote{We discuss a more general parameterization of off-shell momenta in Appendix \ref{sec:mom}.}
\begin{equation}
\begin{split}\label{mompar}
q^A = \o {\hat q}^A , \qquad {\hat q}^A = \left( \frac{ 1 + y^2}{2} \,,\,  y^a \,,\, \frac{ 1 - y^2 }{2} \right) , \qquad \ve^a_A ( {\vec q}\,) =  \p^a {\hat q}_A  =   \left( - y^a , \delta^a_b , - y^a  \right) . 
\end{split}
\end{equation}
Substituting \eqref{mompar} 
into \eqref{gaugefieldmodeexp} and writing the integral over flat null coordinates, we obtain\footnote{Similar mode expansions exist for the other components of the field strength, but we won't need them here.}
\begin{equation}
\begin{split}\label{Furexp}
F^{(R\pm)}_{ur} (u,r,x) = \frac{e}{4(2\pi)^{d+1} r }  \int_0^\infty d\o \int d^d y\,\o^{d-1} \left[  \p^a \CO^\pmm_a(\o,x+y) e^{ - \frac{ i }{2} \o u - \frac{ i }{2} \o r y^2 }  + \cc \right] . 
\end{split}
\end{equation}

To determine the asymptotic expansion of \eqref{Furexp} at large $|u|$ and $|r|$, we assume that the creation and annihilation modes admit a soft expansion of the form
\begin{equation}
\begin{split}\label{softexp}
\p^a \CO^\pmm_a(\o,x+y) = \sum_{n=0}^\infty \o^{n-1} \p^a \CO^{(\pm,n)}_a ( x + y ) \qquad \text{as} \qquad \o \to 0^+.
\end{split}
\end{equation}
This soft expansion allows for a simple pole in $\o$ which is consistent with Weinberg's soft photon theorem. The corresponding ``soft photon operator'' is $\CO_a^{(\pm,0)}(x)$.\footnote{This operator was denoted by $S_a(x)$ in \cite{Kapec:2017gsg}.} It will play a central role in the following section.

We further assume that the coefficients admit a Fourier transform
\begin{equation}
\begin{split}\label{fourierexp}
\p^a \CO^{(\pm,n)}_a ( x + y )  &= \int \frac{d^d k }{ ( 2\pi)^d } e^{i k \cdot ( x + y ) } \SO^\pmm_n ( k )  \\
 \SO^\pmm_n ( k ) &= \int d^d y\, e^{- i k \cdot y} \p^a \CO^{(\pm,n)}_a ( y ) .
\end{split}
\end{equation}
Substituting \eqref{softexp} and \eqref{fourierexp} into \eqref{Furexp}, we find
\begin{equation}
\begin{split}\label{Furterm}
F^{(R\pm)}_{ur} (u,r,x) =  \frac{e}{2(2\pi)^{\frac{d}{2}+1}  }   \sum_{n=0}^\infty \int \frac{d^d k }{ ( 2\pi)^d } k^{2\nu_n}  \left[ \frac{i e^{i k \cdot x }   }{(ir)^{d+n}} \SO^\pmm_n ( k ) (kz)^{-\nu_n} K_{\nu_n} \left( k z \right) + \cc \right] . 
\end{split}
\end{equation}
where
\begin{equation}
\begin{split}\label{zdef}
z = \frac{\sqrt{i u}}{\sqrt{ir}} , \qquad \nu_n = \frac{d}{2}-1+n ,
\end{split}
\end{equation}
and $K_\nu$ is the modified Bessel function of the second kind. We now consider the large $|r|$ expansion of the field strength, or equivalently the small $z$ limit. This takes a qualitatively different form depending on whether $d$ is odd or even, so we consider the two cases separately.

\subsubsection*{Odd Dimensions}

When $d\in2\mzz+1$, the order of the Bessel function is $\nu_n = \frac{d}{2} - 1 + n \not\in \mzz $. For non-integer order, the asymptotic expansion near $z=0$ of the Bessel function is
\begin{equation}
\begin{split}\label{Kexpnonint}
K_\nu(z) = \frac{\pi}{ 2 }  \csc ( \pi \nu )    \sum_{s=0}^\infty  \frac{1}{\G(s+1)}\left[ \frac{ \left(z/2 \right)^{2s-\nu} }{ \G ( 1 + s - \nu ) }  -  \frac{  \left( z/2 \right)^{2s+\nu}   }{  \G ( 1 + s + \nu ) }  \right] , \quad \text{as} \quad z \to 0.
\end{split}
\end{equation}
Substituting \eqref{Kexpnonint} into \eqref{Furterm}, we find
\begin{equation}
\begin{split}\label{Furlargerodd}
F^{(R\pm)}_{ur} (u,r,x) &= \frac{e}{8(2\pi)^{\frac{d}{2}}  }   \sum_{n=0}^\infty   \sum_{s=0}^\infty  \frac{\csc ( \pi \nu_n )  }{ 2^{2s-\nu_n} \G(s+1)\G ( 1 + s - \nu_n ) }  \\
&\qquad \qquad \qquad \qquad \times  \left[ \frac{i (iu)^{s-\nu_n}  }{(ir)^{\frac{d}{2}+1+s}} \int \frac{d^d k }{ ( 2\pi)^d } k^{2s} \SO^\pmm_n ( k ) e^{i k \cdot x }     + \cc \right] \\
&\qquad \qquad - \frac{e}{8(2\pi)^{\frac{d}{2}}  }   \sum_{n=0}^\infty \sum_{s=0}^\infty  \frac{ \csc ( \pi \nu_n )     }{ 2^{2s+\nu_n} \G(s+1) \G ( 1 + s + \nu_n ) } \\
&\qquad\qquad \qquad \qquad \times   \left[ \frac{i (iu)^{s} }{(ir)^{d+n+s}} \int \frac{d^d k }{ ( 2\pi)^d } k^{2(s+\nu_n)}\SO^\pmm_n ( k )  e^{i k \cdot x }   + \cc \right]  . 
\end{split}
\end{equation}
We can now use the inverse Fourier transform \eqref{fourierexp} to go back to the momentum space variables. Evaluating
\begin{equation}
\begin{split}\label{Oint1}
\int \frac{d^d k }{ ( 2\pi)^d } k^{2\k} \SO^\pmm_n ( k ) e^{i k \cdot x }   =     \begin{cases}
( - \p^2 )^\k  \p^a \CO^{(\pm,n)}_a ( x )  & \k \in \mzz  \\
 \frac{ 4^\k \G\left(\frac{d}{2}+\k\right)}{ \pi^{\frac{d}{2}} \G(-\k)} \displaystyle\int d^d y \frac{  \p^a \CO^{(\pm,n)}_a ( y )  }{ \left[  ( x - y  )^2  \right]^{\frac{d}{2} + \k }  } & \k \not\in\mzz
 \end{cases} 
\end{split}
\end{equation}
where $\p^2 = \p^a \p_a$, and substituting into \eqref{Furlargerodd}, we obtain\begin{equation}
\begin{split}\label{Furlargerodd1}
F^{(R\pm)}_{ur} (u,r,x) &= \frac{e}{8(2\pi)^{\frac{d}{2}}  }   \sum_{n=0}^\infty   \sum_{s=0}^\infty  \frac{  2^{\nu_n - 2s} \csc ( \pi \nu_n )  }{ \G(s+1)\G ( 1 + s - \nu_n ) } \\
&\qquad \qquad \qquad \qquad \qquad  \times \left[ \frac{i (iu)^{s-\nu_n}  }{(ir)^{\frac{d}{2}+1+s}}  ( - \p^2 )^s  \p^a \CO^{(\pm,n)}_a ( x )   + \cc \right] \\
&\qquad \qquad \qquad + \frac{e}{2^{\frac{d}{2}+3}\pi^{d+1}  }   \sum_{n=0}^\infty \sum_{s=0}^\infty     (-1)^{s}   \frac{ \G\left( \frac{d}{2}+\nu_n+s \right)  }{ 2^{-\nu_n} \G(s+1)  }  \\
&\qquad \qquad \qquad  \qquad \qquad \times  \left[ \frac{i (iu)^{s} }{(ir)^{d+n+s}}   \int d^d y \frac{  \p^a \CO^{(\pm,n)}_a ( y )  }{\left[  ( x - y  )^2  \right]^{\frac{d}{2} + \nu_n + s }  }  + \cc \right]  . 
\end{split}
\end{equation}

\subsubsection*{Even Dimensions}

When $d\in2\mzz$, the order of the Bessel function is $\nu_n = \frac{d}{2} - 1 + n \in \mzz$. For integer order,  the asymptotic expansion near $z=0$ of the Bessel function is given by
\begin{equation}
\begin{split}\label{Kexpint}
K_\nu(z) &= \sum_{s=0}^{\nu-1} \frac{ ( - 1 )^s \G(\nu-s) \left( \frac{z}{2} \right)^{2s-\nu}  }{ 2 \G(s+1) } \\
&\qquad \qquad \qquad \qquad \qquad + \sum_{s=0}^\infty \frac{ (-1)^{\nu-1} \left[  \log \left(  \frac{z}{2}  e^{\g_E} \right) - \frac{ 1 }{ 2  } \left( H_s + H_{s+\nu}  \right)  \right] \left( \frac{z}{2} \right)^{2s+\nu} }{ \G(s+1) \G(s+\nu+1) } ,
\end{split}
\end{equation}
where $\gamma_E$ is the Euler-Mascheroni constant and $H_n$
is the $n$-th harmonic number (we define $H_0=0$). Substituting \eqref{Kexpint} into \eqref{Furterm}, we find
\begin{equation}
\begin{split}\label{Furlargereven}
&F^{(R\pm)}_{ur} (u,r,x) \\
&\quad =\frac{e}{2(2\pi)^{\frac{d}{2}+1}  }   \sum_{n=0}^\infty \sum_{s=0}^{\nu_n-1} \frac{ ( - 1 )^s \G(\nu_n-s)  }{ 2^{2s-\nu_n+1} \G(s+1) }  \left[ \frac{i ( iu)^{s-\nu_n}   }{(ir)^{\frac{d}{2}+1+s}} \int \frac{d^d k }{ ( 2\pi)^d } k^{2s} \SO^\pmm_n ( k ) e^{i k \cdot x }   + \cc \right] \\
&\qquad \qquad \qquad + \frac{e}{2(2\pi)^{\frac{d}{2}+1}  }   \sum_{n=0}^\infty \sum_{s=0}^\infty \frac{  2^{-2s-\nu_n} (-1)^{\nu_n-1} }{\G(s+1) \G(s+\nu_n+1) } \int \frac{d^d k }{ ( 2\pi)^d } k^{2\left(s+\nu_n\right)}  \\
&\qquad \qquad \qquad \quad \times  \left[ \frac{i (iu)^s  \left[  \log \left(  \frac{k}{2} \frac{ \sqrt{iu} }{ \sqrt{ir}}  e^{\g_E} \right) - \frac{ 1 }{ 2  } \left( H_s + H_{s+\nu_n}  \right)  \right] }{(ir)^{d+n+s}} \SO^\pmm_n ( k ) e^{i k \cdot x }     + \cc \right] .   
\end{split}
\end{equation}
Again, we want to use the inverse Fourier transform \eqref{fourierexp} to go back to the momentum space variables. Evaluating for $\kappa \in \mzz$
\begin{equation}
\begin{split}\label{Oint2}
	\int \frac{d^d k }{ ( 2\pi)^d } k^{2\k} \log k \, \SO^\pmm_n ( k )  e^{i k \cdot x }  &= \frac{(4\pi)^{\frac{d}{2}}   \G (\k+1)   \G \left(\frac{d}{2}+\k\right)  }{ (-1)^{\k+1} 2^{1-2 \k} } \int d^d y  \frac{\p^a \CO^{(\pm,n)}_a ( y ) }{\left[(x-y)^2\right]^{\frac{d}{2}+\k} }  ,
\end{split}
\end{equation}
we get upon substituting \eqref{Oint1} and \eqref{Oint2} into \eqref{Furlargereven}
\begin{equation}
\begin{split}\label{Furlargereven1}
& F^{(R\pm)}_{ur} (u,r,x) \\
&= \frac{e}{2(2\pi)^{\frac{d}{2}+1}  }   \sum_{n=0}^\infty \sum_{s=0}^{\nu_n-1} \frac{ ( - 1 )^s \G(\nu_n-s)  }{ 2^{2s-\nu_n+1} \G(s+1) }  \left[ \frac{i ( iu)^{s-\nu_n}   }{(ir)^{\frac{d}{2}+1+s}}   ( - \p^2 )^s  \p^a \CO^{(\pm,n)}_a ( x )   + \cc \right] \\
&~ + \frac{e}{8\pi}   \sum_{n=0}^\infty \sum_{s=0}^\infty \frac{(-1)^{s}  2^{\frac{d}{2} + \nu_n} \G \left(\frac{d}{2}+s+\nu_n\right)   }{\G(s+1)  }  \left[ \frac{i (iu)^s    }{(ir)^{d+n+s}}   \int d^d y  \frac{\p^a \CO^{(\pm,n)}_a ( y ) }{  \left[(x-y)^2\right]^{\frac{d}{2}+\nu_n+s} }    + \cc \right]\\
&~ + \frac{e}{2(2\pi)^{\frac{d}{2}+1}  }   \sum_{n=0}^\infty \sum_{s=0}^\infty \frac{  2^{-2s-\nu_n} (-1)^{\nu_n-1} }{\G(s+1) \G(s+\nu_n+1) }   \\
&\qquad \times  \left[ \frac{i (iu)^s  \left[  \log \left(  \frac{1}{2} \frac{ \sqrt{iu} }{ \sqrt{ir}}  e^{\g_E} \right) - \frac{ 1 }{ 2  } \left( H_s + H_{s+\nu_n}  \right)  \right] }{(ir)^{d+n+s}}  ( - \p^2 )^{s+\nu_n} \p^a \CO^{(\pm,n)}_a ( x )  + \cc \right]  .   
\end{split}
\end{equation}

\subsubsection{Coulombic Field}

Having performed the asymptotic expansion for the radiative field, we now turn to the Coulombic field. The large $r$ expansion for this can be performed in two ways. First, we could follow strategy used in the previous section and start by writing the explicit solution for the Coulombic field using the massless scalar Green's function as
\begin{equation}
\begin{split}\label{greenfuncsol}
F_{AB}^{(C\pm)} (X) = e^2 \int d^{d+2} Y {\mathfrak G}^\pmm(X-Y) \left[ \p_A J_B (Y) - \p_B J_A(Y) \right] , 
\end{split}
\end{equation}
where ${\mathfrak G}^\pmm(X)$ is the retarded $(-)$ or advanced $(+)$ massless scalar Green's function satisfying $ \p^2 {\mathfrak G}^\pmm(X) = \delta^{(d+2)}(X)$. Explicitly,
\begin{equation}
\begin{split}\label{greenfunc}
{\mathfrak G}^\pmm(X) =  2 \theta(\mp X^0) \text{Re}\,[f_d(X)] , \qquad f_d(X) = \frac{ ( - i )^{d+1} \Gamma \left( \frac{d}{2} \right)  }{4\pi^{\frac{d+2}{2}}  \left[ (X^0)^2 - | \vec{X}\,|^2 - i \e \right]^{\frac{d}{2}} }  . 
\end{split}
\end{equation}
Then, assuming an appropriate asymptotic expansion (near $\ci^\pm$) for the conserved current (along with some other assumptions regarding the behaviour of the current near $i^\pm$), we can use \eqref{greenfuncsol} and \eqref{greenfunc} to determine the large $r$ expansion of the Coulombic field. 

While possible, this is a tedious exercise and it is more convenient to determine the large $r$ expansion of the Coulombic field as follows. We start by assuming that the Coulombic field admits a Taylor expansion of the following form near $\ci^\pm$ (consistent with \eqref{Ffalloff}):
\begin{equation}
\begin{split}\label{FCexp}
F_{ur}^{(C\pm)} (u,r,x) &= \sum_{n=0}^\infty \frac{ F_{ur}^{(C\pm,d+n)} (u,x) }{ |r|^{d+n} } , \qquad F_{ua}^{(C\pm)} (u,r,x) = \sum_{n=0}^\infty \frac{ F_{ua}^{(C\pm,d-1+n)} (u,x) }{ |r|^{d-1+n} }  \\
F_{ra}^{(C\pm)} (u,r,x) &= \sum_{n=0}^\infty \frac{ F_{ra}^{(C\pm,d+n)} (u,x) }{ |r|^{d+n} } , \qquad F_{ab}^{(C\pm)} (u,r,x) = \sum_{n=0}^\infty \frac{ F_{ab}^{(C\pm,d-1+n)} (u,x) }{ |r|^{d-1+n} } .
\end{split}
\end{equation}
Similarly, consistent with \eqref{currfalloff}, we also assume the expansion
\begin{equation}
\begin{split}\label{Jexp}
J_u (u,r,x) &= \sum_{n=0}^\infty \frac{ J_u^{(\pm,d+n)} (u,x) }{ |r|^{d+n} }  \\
J_a (u,r,x) &= \sum_{n=0}^\infty \frac{ J_a^{(\pm,d+n)} (u,x) }{ |r|^{d+n} } \\
J_r (u,r,x) &= \sum_{n=0}^\infty \frac{ J_r^{(\pm,d+2+n)} (u,x) }{ |r|^{d+2+n} } .
\end{split}
\end{equation}
Substituting \eqref{FCexp} and \eqref{Jexp} into \eqref{MaxwellEq}, we can obtain equations order by order in the large $r$ expansion. Solving these yields the full asymptotic expansion for the Coulombic field strength, though for our purposes, it suffices to obtain just the following leading order constraint equation (obtained from the $u$ component of Maxwell's equations):
\begin{equation}
\begin{split}\label{cons1}
2 \p_u F^{(C\pm,d)}_{ur}  &= e^2 J^{(\pm,d)}_u .  
\end{split}
\end{equation}

\section{Weinberg's Soft Photon Theorem}\label{sec:softphotonthm}

\subsection{Matching Condition}

As in four dimensions, we impose a matching condition on the radial electric field\footnote{The unusual sign is due to the coordinate choice. To see that this is correct, one can verify it for the Li\'enard-Wiechert potential for a static charge, given by $A_0 (\vec{X}) = c |\vec{X}|^{1-d} $ and $\vec{A}(\vec{X}) = 0$. Moving to flat null coordinates, we find
$$
F_{ur} = \frac{c}{2} \left[  \big(1+x^2\big)\p_u - \p_r \right] \big|\vec{X}\big|^{1-d} = \frac{2^{d-1} c ( d - 1 )  ( r  + rx^2 - u )} { \left[ u^2 - 2 u r ( 1 - x^2 ) + r^2 ( 1 + x^2 )^2 \right]^{\frac{d+1}{2}}}.
$$
This implies $\lim\limits_{r \to \pm \infty} \big( |r|^d F_{ur} \big) = \pm 2^{d-1} c (d-1) (1+x^2)^{-d}$. For more extensive discussions on the antipodal matching condition, we refer the reader to \cite{Campiglia:2017mua,Prabhu:2018gzs}.}
\begin{equation}
\begin{split}\label{matchingcond}
\left.F_{ur}^{(+,d)} \right|_{\ci^+_-}  &= - \left.F_{ur}^{(-,d)} \right|_{\ci^-_+} .
\end{split}
\end{equation}
Recall that coordinate $x$ labels antipodal points on the celestial spheres on $\ci^+$ and $\ci^-$, so the above condition is an \emph{antipodal} matching condition. To massage this matching condition into a more useful form, we define the ``charge''
\begin{equation}\label{charge}
\begin{split}
	Q^\pm_\ve = \pm \frac{2}{e^2} \int_{\ci^\pm_\mp} d^d x\, \ve F_{ur}^{(\pm,d)}, 
\end{split}
\end{equation}
where $\ve \equiv \ve(x)$ is a function defined on the $d$-dimensional celestial sphere at $\ci^\pm$. As we will show in \S\ref{sec:largegaugesym} using the covariant phase space formalism, $Q^\pm_\ve$ is the charge that generates large gauge transformations on $\ci^\pm$. We can therefore think of $Q^\pm_\ve$ as measuring the local $U(1)$ charge of the $in$ and $out$ state respectively. For $\ve = 1$, this measures the total global $U(1)$ charge. These charges can be determined in experiment by measuring the electromagnetic memory \cite{Yoshida:2017fao,Pasterski:2015zua,Susskind:2015hpa}.

The antipodal matching condition \eqref{matchingcond} immediately implies
\begin{align}\label{matchcond1}
	Q^+_\ve = Q^-_\ve .
\end{align}
Breaking up the field strength into radiative and Coulombic pieces, we can write the charges as
\begin{equation}\label{chargedecomp}
\begin{split}
Q^\pm_\ve = Q^{\pm S}_\ve   +  Q^{\pm H}_\ve ,
\end{split}
\end{equation}
where
\begin{equation}
\begin{split}
Q^{\pm S}_\ve = \pm\frac{2}{e^2} \int_{\ci^\pm_\mp} d^d x\, \ve F_{ur}^{(R\pm,d)}  , \qquad Q^{\pm H}_\ve = \pm\frac{2}{e^2} \int_{\ci^\pm_\mp} d^d x \,\ve F_{ur}^{(C\pm,d)}   .
\end{split}
\end{equation}
$Q^{\pm S}_\ve$ are the incoming and outgoing soft charges and $Q^{\pm H}_\ve$ are the incoming and outgoing hard charges. Using \eqref{cons1}, the hard charges can be written as
\begin{equation}\label{hardmassive}
\begin{split}
Q^{\pm H}_\ve &=  - \int_{\ci^\pm} du d^d x\, \ve  J^{(\pm,d)}_u \pm \frac{2}{e^2} \int_{\ci^\pm_\pm}  d^d x\, \ve F_{ur}^{(C\pm,d)}  . 
\end{split}
\end{equation}
The second term above is the Coulombic field in the far past or far future and only receives contributions from stable massive particles. In this paper, we assume that our theory contains only stable massless states so this contribution vanishes. Thus, the hard charges become
\begin{equation}
\begin{split}\label{hardcharge}
Q^{\pm H}_\ve &=  - \int_{\ci^\pm} du d^d x \,\ve  J^{(\pm,d)}_u  . 
\end{split}
\end{equation}

To determine the soft charge in odd and even dimensions, we extract the coefficient of $|r|^{-d}$ in \eqref{Furlargerodd1} and \eqref{Furlargereven1}, respectively. In odd dimensions, we get
\begin{equation}
\begin{split}\label{Furdodd}
F^{(R\pm,d)}_{ur}  &= \pm \frac{e(-1)^{\frac{d-1}{2}}  \G\left( d-1 \right) }{ 16 \pi^{d+1}  }  \int d^d y \frac{  \p^a \CO^{(\pm,0)}_a ( y ) +  \p^a \CO^{(\pm,0)\dagger}_a ( y )  }{ \left[  ( x - y  )^2  \right]^{ d - 1  }  }  ,
\end{split}
\end{equation}
and for even dimensions, we get
\begin{equation}
\begin{split}\label{Furdeven}
F^{(R\pm,d)}_{ur} &= - \frac{e}{  (4\pi)^{\frac{d}{2}+1} \G\left(\frac{d}{2}\right)  }   \sum_{n=1}^\infty  2^n \G(n) \left[ i ( iu)^{-n}   ( - \p^2 )^{\frac{d}{2}-1}  \p^a \CO^{(\pm,n)}_a ( x )   + \cc \right] \\
&\qquad \qquad + \frac{ ie  (-1)^{\frac{d}{2}} \G \left(d-1\right)   }{ 2^{4-d}\pi}  \int d^d y  \frac{ \p^a \CO^{(\pm,0)}_a ( y ) - \p^a \CO^{(\pm,0)\dagger}_a ( y ) }{  \left[(x-y)^2\right]^{d-1} } \\
&\qquad \qquad + \frac{ i e\left[ \log \left(  \frac{ |u| e^{2\g_E}  }{ 4 |r| }   \right)  -  H_{\frac{d}{2}-1} \right] }{(4\pi)^{\frac{d}{2}+1}\G\left(\frac{d}{2}\right)  }     ( - \p^2 )^{\frac{d}{2}-1}  \left[ \p^a \CO^{(\pm,0)}_a ( x )  - \p^a \CO^{(\pm,0)\dagger}_a ( x ) \right] \\
&\qquad \qquad - \frac{e\left[ \T(u) - \T(r) \right] }{8(4\pi)^{\frac{d}{2}}\G\left(\frac{d}{2}\right)  }      ( - \p^2 )^{\frac{d}{2}-1} \left[   \p^a \CO^{(\pm,0)}_a ( x )  +  \p^a \CO^{(\pm,0)\dagger}_a ( x )   \right] .   
\end{split}
\end{equation}
Notice that in even dimensions this coefficient has a logarithmic divergence, which means the soft charge is potentially ill-defined in even dimensions. To cancel this divergence, we impose\footnote{Note that \eqref{modesconstraint} also implies that we can no longer think of $\CO^{(\pm,0)\dagger}_a$ as a creation operator. Rather, it is an operator that shifts the vacuum. From the perspective of the path integral, this condition is equivalent to continuity of the $S$-matrix at $\o = 0$, i.e. the zero energy limit of the operators does not depend on whether we approach zero from the positive or negative real axis (see \cite{He:2015zea}).} 
\begin{equation}
\begin{split}\label{modesconstraint}
  \p^a \CO^{(\pm,0)}_a ( x )  =  \p^a \CO^{(\pm,0)\dagger}_a ( x ) . 
\end{split}
\end{equation}
We remark that this constraint implies that in even dimensions $\left.F_{ur}^{(R\pm,d)}\right|_{\ci^\pm_\pm} = 0$, an assumption that was made previously in \cite{He:2014cra}. The soft charge now takes the form
\begin{equation}\label{softcharges}
\begin{split}
Q^{\pm S}_\ve = \begin{cases}
 \frac{1}{(4\pi)^{\frac{d}{2}}\G\left(\frac{d}{2}\right) e }   \displaystyle\int d^d x \,\ve(x)     ( - \p^2 )^{\frac{d}{2}-1}  \p^a \CO^{(\pm,0)}_a ( x )   & d \in 2 \mzz  \\
\frac{(-1)^{\frac{d-1}{2}}  \G\left( d-1 \right) }{ 4 \pi^{d+1}  e  }   \displaystyle\int d^d x \,\ve(x)  \displaystyle\int d^d y \frac{  \p^a \CO^{(\pm,0)}_a ( y ) }{ \left[  ( x - y  )^2  \right]^{ d - 1  }  }  & d \in 2 \mzz + 1 . 
\end{cases}
\end{split}
\end{equation}

We conclude this section by bringing the soft charge into the same form for both odd and even dimensions by judiciously choosing $\ve$ for the remainder of the paper to be
\begin{equation}\label{fixve}
\begin{split}
	\ve(z) = f_x(z) = ( - \p^2 ) \log \left[ ( x - z )^2  \right] .
\end{split}
\end{equation}
Then, using the fact that in even dimensions
\begin{equation}
\begin{split}
( - \p^2 )^{\frac{d}{2}-1} f_x(z)  &= ( - \p^2 )^{\frac{d}{2}} \log \left[ ( x - z )^2  \right]  = - ( 4\pi )^{\frac{d}{2}} \Gamma \left( \frac{d}{2} \right) \delta^{(d)}(x-z) , 
\end{split}
\end{equation}
and in odd dimensions
\begin{equation}
\begin{split}
	 \int d^d z \frac{ f_x(z)  }{\left [  ( z - y  )^2  \right]^{ d - 1  }  }  &= -2(d-2) \int d^d z \frac{ 1 }{ \left[  ( z - y  )^2  \right]^{ d - 1  } ( x - z )^2  }  \\
	 &= -\frac{4 (-1)^{\frac{d-1}{2}} \pi^{d+1} }{\Gamma (d-1)}  \delta^{(d)}(x-y),
\end{split}
\end{equation}
the soft charge becomes for all spacetime dimensions\footnote{Although we picked $\ve = f_x$ in \eqref{fixve}, we can recover the soft charge $Q_\ve^{\pm S}$ for general $\ve$ from \eqref{softall} by using \eqref{softcharges} to write $Q_\ve^{\pm S}$ as an appropriate linear combination of $Q^{\pm S}_{f_x} \sim \p^a \CO_a^{(\pm,0)}$. This means we didn't lose any generality by choosing $\ve = f_x$, and the Ward identity we derive below is also completely general.}
\begin{align}\label{softall}
	Q_{f_x}^{\pm S} = -\frac{1}{2e}\left(\p^a\CO_a^{(\pm,0)}(x) + \p^a\CO_a^{(\pm,0)\dag}(x)\right) = -\frac{1}{e}\p^a\CO_a^{(\pm,0)}(x).
\end{align}

\subsection{Ward Identity}

Thus far, our discussion of fields, matching conditions, and charges has been entirely classical. We now extend the discussion to a semi-classical theory and derive a Ward identity. In such a theory, the quantity of interest is the scattering amplitude, which is the overlap between the $in$ state (living on $i^-\cup\ci^-$) and the $out$ state (living on $i^+\cup \ci^+$), i.e.
\begin{equation}
\begin{split}
	\CA_n = \braket{\text{out}}{\text{in}},
\end{split}
\end{equation}
where $n$ is the total number of particles in the $in$ and $out$ state. The classical matching condition \eqref{matchcond1} then translates to
\begin{equation}
\begin{split}\label{wardid1}
\bra{\text{out}} \left( Q^+_\ve - Q^-_\ve \right) \ket{\text{in}} = 0 . 
\end{split}
\end{equation}
Using \eqref{chargedecomp}, we can rewrite this as
\begin{equation}
\begin{split}\label{wardid2}
\bra{\text{out}} \left( Q^{+S}_\ve    -    Q^{-S}_\ve \right)\ket{\text{in}} = - \bra{\text{out}}   \left( Q^{+H}_\ve - Q^{-H}_\ve  \right) \ket{\text{in}} .
\end{split}
\end{equation}

To simplify this, we evaluate the action of the hard charge on one-particle states, i.e. the right-hand-side of \eqref{wardid2}. Let $\ket{\Psi_i,\vec{p}_i,s_i}$ be a massless one-particle state with charge $Q_i$, momentum $\vec{p}_i$, and spin $s_i$. We may parameterize the momentum as
\begin{equation}
\begin{split}
p_i^A = \o_i \left(  \frac{ 1 + x_i^2 }{ 2 }  , x_i^a , \frac{ 1 - x_i^2 }{ 2 } \right) . 
\end{split}
\end{equation}
The action of the conserved current on the bra and ket states is (see Appendix \ref{app:scalar} for an explicit calculation for scalar fields)
\begin{equation}
\begin{split}
\int_{\ci^-} du J_u^{(-,d)}(u,x)  \ket{\Psi_i,\vec{p}_i,s_i} &=  \ket{\Psi_i,\vec{p}_i,s_i} Q_i \delta^{(d)} \left( x - x_i \right)  \\
\bra{\Psi_i,\vec{p}_i,s_i}  \int_{\ci^+} du J_u^{(+,d)}(u,x) &=  Q_i \delta^{(d)} \left( x - x_i \right) \ket{\Psi_i,\vec{p}_i,s_i}  . \\
\end{split}
\end{equation}
It follows using \eqref{hardcharge} that
\begin{equation}
\begin{split}\label{Qhardaction}
Q^{-H}_\ve\ket{\Psi_i,\vec{p}_i,s_i} =   -  Q_i \ve(x_i)   \ket{\Psi_i,\vec{p}_i,s_i} , \qquad \bra{\Psi_i,\vec{p}_i} Q^{+H}_\ve =    - Q_i \ve(x_i) \bra{\Psi_i,\vec{p}_i,s_i}  . \\
\end{split}
\end{equation}
Note that this is simply an infinitesimal large gauge transformation of the state. Setting $\ve = f_x$ in \eqref{wardid2} and using \eqref{softall} and \eqref{Qhardaction}, we derive the Ward identity
\begin{equation}
\begin{split}
\label{mainwardid}
&\bra{\text{out}}  \left[  \p^a \CO^{(+,0)}_a (x)    -    \p^a \CO^{(-,0)}_a (x)  \right] \ket{\text{in}} = e  \sum_{i=1}^n \eta_i Q_i\p^2 \log \left[ ( x - x_i )^2 \right]   \braket{\text{out}}{\text{in}} .
\end{split}
\end{equation}
Here, $\eta_i = +1$ for outgoing particles and $\eta_i = -1$ for incoming particles.

\subsection{Soft Theorem}

Finally, in this section we will show that the Ward identity \eqref{mainwardid} is implied by Weinberg's leading soft photon theorem, which in standard momentum space variables reads
\begin{equation}\label{softthmmom}
\begin{split}
\CA^{\text{out}}_{n+1}(\vec{p}_\g,\ve_a ; p_1 , \cdots , p_n ) = e \sum_{i=1}^n \eta_i Q_i \frac{ p_i \cdot \ve_a(p_\g) }{ p_i \cdot p_\g } \CA_n ( p_1 , \cdots , p_n  )  + O \big( (p_\g)^0 \big) ,
\end{split}
\end{equation}
where $\ve_a$ is the polarization vector (not to be confused with the function $\ve(z)$ in \eqref{fixve}). Note that the leading order term in the soft expansion is a simple pole in photon momentum $p_\g$, and this is known as the Weinberg pole.\footnote{The first subleading term in the soft expansion is also universal and is related to an asymptotic symmetry, but this is beyond the scope of this paper.} We want to parametrize \eqref{softthmmom} using the momentum space parameterization employed in the previous sections (see \S\ref{sec:mom} for details), i.e.
\begin{equation}
\begin{split}\label{mompolpar}
	p_\g^A &= \o \left( \frac{1+x^2}{2} , x^a , \frac{1-x^2}{2} \right)  \\
	 p_i^A &=  \o_i \left( \frac{ 1 + x_i^2 }{2} , x_i^a , \frac{ 1 - x_i^2 }{ 2 } \right)   \\
	\ve^A_a(p_\g) &= \left( x_a , \delta^b_a , - x_a \right) . 
\end{split}
\end{equation}
In this parametrization, the soft factor corresponding to polarization $a$ is 
\begin{equation}\label{softrhs}
\begin{split}
e \sum_{i=1}^n \eta_i Q_i \frac{ p_i \cdot \ve_a(p_\g) }{ p_i \cdot p_\g }  = \frac{e}{\omega}  \sum_{i=1}^n \eta_i Q_i \p_a \log \left[ ( x - x_i )^2 \right] .
\end{split}
\end{equation}

The last step is to recast the scattering amplitudes in \eqref{softthmmom} in terms of the $in$ and $out$ states. The $n$-point amplitude on the right-hand-side is simply $\braket{\text{out}}{\text{in}}$. On the other hand, an outgoing photon corresponds to the insertion of the operator $\CO^\+_a(\o,x) - \CO^{\-}_a(\o,x)$ (see Appendix \ref{softinsertion} for details). The coefficient of $\o^{-1}$ in \eqref{softrhs} is extracted by first multiplying by $\o$ and then taking a $\o \to 0$ limit. Using \eqref{softexp}, we find that this operator is
\begin{align}
	\lim_{\o \to 0}  \left[ \o\CO^\+_a(\o,x) - \o\CO^{\-}_a(\o,x) \right] = \CO_a^{(+,0)}(x) -  \CO_a^{(-,0)}(x) .
\end{align}
Substituting this into an $S$-matrix element gives precisely the left-hand-side of \eqref{softthmmom}, and using \eqref{softrhs}, we see that the soft theorem can be written as
\begin{equation}
\begin{split}\label{softthmfinal}
\bra{\text{out}} \left( \CO_a^{(+,0)}(x)   -    \CO_a^{(-,0)}(x) \right) \ket{\text{in}} =  e \sum_{i=1}^n \eta_i Q_i \p_a \log \left[ ( x - x_i )^2 \right] \braket{\text{out}}{\text{in}}.
\end{split}
\end{equation}
Acting on both sides with $\p^a$ and using \eqref{modesconstraint}, we immediately reproduce the Ward identity \eqref{mainwardid}, completing our demonstration that Weinberg's leading soft photon theorem implies the Ward identity.

We conclude this section with a few comments.

\begin{itemize}
\item We have shown above that Weinberg's leading soft photon theorem for an outgoing soft photon implies the Ward identity. However, we could have just as well chosen to work with an incoming soft photon, in which case the soft theorem reads
\begin{equation}\label{softthmmom1}
\begin{split}
\CA^{\text{in}}_{n+1}(\vec{p}_\g,\ve_a ; p_1 , \cdots , p_n ) = - e \sum_{i=1}^n \eta_i Q_i \frac{ p_i \cdot \ve_a(p_\g) }{ p_i \cdot p_\g } \CA_n ( p_1 , \cdots , p_n  )  + O \big( (p_\g)^0 \big),
\end{split}
\end{equation}
and differs from \eqref{softthmmom} by only a relative sign. The left-hand-side of the above equation corresponds to the insertion of $\CO_a^{(-)\dagger}(\o,x) - \CO_a^{(+)\dagger}(\o,x)$. Multiplying by $\o$, taking the $\o \to 0$ limit, and using \eqref{modesconstraint}, we immediately find that the outgoing soft photon theorem \eqref{softthmmom} \emph{implies} the incoming soft photon theorem, and vice versa.

\item We have not yet demonstrated the equivalence between the Ward identity and Weinberg's leading soft photon theorem. Instead, we have only established the fact that the soft theorem implies the Ward identity. To prove the converse, we need to use the additional constraint that
\begin{equation}
\begin{split}\label{addtnlconstraint}
\p_a \CO_b^{(\pm,0)}(x)  - \p_b \CO_a^{(\pm,0)}(x) = 0  .
\end{split}
\end{equation}
This constraint is obviously true from the standpoint of the soft theorem \eqref{softthmfinal},\footnote{This can be derived from a careful study of the symplectic form, which will be done in detail in forthcoming work. This constraint was also discussed in \cite{Strominger:2017zoo}.} and is solved by
\begin{equation}
\begin{split}\label{Sdefff}
\CO_a^{(\pm,0)}(x)  = \p_a \CS^\pmm (x)  . 
\end{split}
\end{equation}
We can then derive the Ward identity for the insertion of $\CS^\pmm(x)$, which is given by\footnote{To do this, we choose $\ve(z) = \log\left[(x-z)^2\right]$ in the Ward identity \eqref{wardid2}.} 
\begin{equation}
\begin{split}
&\bra{\text{out}}  \left[ \CS^\+ (x) -  \CS^\- (x)  \right] \ket{\text{in}} = e  \sum_{i=1}^n \eta_i Q_i  \log \left[ ( x - x_i )^2 \right]   \braket{\text{out}}{\text{in}} .
\end{split}
\end{equation}
Acting on both sides with $\p_a$ and using \eqref{Sdefff}, we immediately reproduce the soft theorem \eqref{softthmfinal}.

\end{itemize}

\section{Large Gauge Symmetry}\label{sec:largegaugesym}

In the previous section, we have shown that the Ward identity corresponding to the charge $Q^\pm_\ve$ is equivalent Weinberg's soft photon theorem. We now show that this charge generates \emph{large gauge transformations} on $\ci^\pm$ using the covariant phase space formalism \cite{Wald:1999wa}.\footnote{A recent analysis of the charge generating large gauge transformations at \emph{spatial infinity} in all dimensions $d\geq 4$ is given in \cite{Esmaeili:2019hom}.} We have already done this for the hard charge, so it remains to show that the soft charge (i.e. the radiative field) acts on the gauge field to generate large gauge transformations. For this purpose, we can focus simply on pure $U(1)$ gauge theory. This is described by the action
\begin{equation}
\begin{split}\label{U1action}
S[A] = - \frac{1}{2e^2} \int_M F \wedge \ast F  , \qquad F = d A , \qquad A \sim A + d \ve . 
\end{split}
\end{equation}
Varying this, we find
\begin{equation}
\begin{split}
\delta S =  - \frac{1}{e^2} \int_M \delta A \wedge ( d \ast F ) - \frac{1}{e^2}   \int_{\p M}  \delta A \wedge \ast F    . 
\end{split}
\end{equation}
The bulk term gives us the equations of motion $d \ast F = 0$. From the boundary term we can read off the canonical one-form on a Cauchy surface $\Sigma$ as
\begin{equation}
\begin{split}
\Theta_\Sigma (\delta) &=  \frac{1}{e^2} \int_\Sigma  \delta A \wedge \ast F. 
\end{split}
\end{equation}
If we restrict $A$ to satisfy the equations of motion and $\delta A$ to satisfy the linearized equations of motion, then $\Theta_\Sigma$ can be understood as a one-form at the point $A$ in the phase space $\Gamma$. The presymplectic form on $\Sigma$ is
\begin{equation}
\begin{split}
\Omega_\Sigma(\delta,\delta') &= \delta \Theta_\Sigma (\delta' )  - \delta' \Theta_\Sigma (\delta) =   \frac{1}{e^2} \int_\Sigma \left[ \delta' A \wedge \ast \delta F - \delta A \wedge \ast \delta' F \right] .
\end{split}
\end{equation}

At this point, we recall that the gauge field $A$ is determined only up to a gauge choice \eqref{U1action}. To keep track of this, we separate the gauge field into two pieces
\begin{equation}
\begin{split}
A = {\tilde A} + d \psi ,
\end{split}
\end{equation}
where ${\tilde A}$ satisfies a chosen gauge-fixing condition,\footnote{In our work, it is often convenient to choose $f[{\tilde A}]={\tilde A}_u$.}
\begin{equation}
\begin{split}
f [ {\tilde A} ]  = 0 ,
\end{split}
\end{equation}
and gauge transformations simply involve
\begin{equation}
\begin{split}
\psi \to \psi + \ve . 
\end{split}
\end{equation}
It follows that the symplectic form is
\begin{equation}
\begin{split}
\Omega_\Sigma(\delta,\delta') &=  \frac{1}{e^2} \int_\Sigma \left[ \delta' {\tilde A} \wedge \ast \delta {\tilde F} - \delta {\tilde A} \wedge \ast \delta' {\tilde F} \right]  + \frac{1}{e^2} \int_{\p\Sigma} \left[ \delta' \psi   \ast \delta F - \delta \psi   \ast \delta' F \right]  . 
\end{split}
\end{equation}
We note that modes for which $\psi |_{\p \Sigma} = 0$ cannot appear in the symplectic form and need to be fixed in order to have a non-degenerate symplectic form. We assume that this has been done, and so the remaining boundary value of $\psi$ symplectically couples to the boundary field strength.

The charge that generates a particular transformation $\Delta$ on the phase space satisfies
\begin{equation}
\begin{split}
\delta Q^\Sigma_\Delta = \O_\Sigma ( \delta , \Delta ) .
\end{split}
\end{equation}
In particular, for gauge transformations $\delta_\ve {\tilde A } = \delta_\ve F = 0$ and $\delta_\ve \psi = \ve$, we get
\begin{equation}
\begin{split}
Q^\Sigma_\ve  =  \frac{1}{e^2} \int_{\p\Sigma} \ve \ast F . 
\end{split}
\end{equation}
Note that the charge is non-zero only if $\ve|_{\p\Sigma} \neq 0$, so this means $Q^\Sigma_\ve$ is the generator of \emph{large gauge transformations}. To determine its form on $\ci^\pm$, we note that the directed area element on $\ci_\mp^\pm$ is 
\begin{equation}
\begin{split}
d S^\pm_{\mu\nu} = \pm  d^d x |r|^d \delta^r_{[\mu} \delta^u_{\nu]}  . 
\end{split}
\end{equation}
The $\pm$ sign arises from the outward-pointing null vector, which on $\ci^\pm$ is $\mp \p_\mu r$. Thus, we have
\begin{equation}
\begin{split}
Q^\pm_\ve  =  \pm \frac{2}{e^2} \int_{\ci^\pm_\mp} d^d x\, \ve F^{(\pm,d)}_{ur} ,
\end{split}
\end{equation}
which is precisely the charge \eqref{charge}.

\appendix

\section*{Acknowledgements}
We would like to thank Eduardo Casali, Indranil Halder, Daniel Kapec, Alok Laddha, Monica Pate, and Kartik Prabhu for useful conversations, and especially Andrew Strominger for reading a preprint and offering helpful comments. TH is grateful to be supported by U.S. Department of Energy grant DE-SC0009999 and by funds from the University of California. PM gratefully acknowledges support from U.S. Department of Energy grant DE-SC0009988.

\section{Coordinate Conventions}

\subsection{Position Space}\label{sec:coordinates}

In this subsection, we will gain some intuition regarding the flat null coordinates utilized throughout this paper by collecting miscellaneous facts about them. Flat null coordinates $x^\mu = (u,x^a,r)$ are related to Cartesian coordinates $X^A$ via
\begin{equation}
\begin{split}
X^0 = \frac{r}{2} \left( 1 + x^2 \right) + \frac{u}{2} , \qquad X^a = r x^a , \qquad X^{d+1} = \frac{r}{2} \left( 1 - x^2 \right) - \frac{u}{2} .
\end{split}
\end{equation}
The inverse relation is
\begin{equation}
\begin{split}
u = \frac{ \big(X^0\big)^2 - X^a X_a - \big( X^{d+1}\big)^2 }{ X^0 + X^{d+1} } , \qquad x^a = \frac{X^a}{X^0 + X^{d+1}} , \qquad  r = X^0 + X^{d+1} . 
\end{split}
\end{equation}
Though the coordinates are ill-defined on the hypersurface $X^0 + X^{d+1} = 0$ where $r  = 0$, away from this point, the metric is well defined and takes the form
\begin{equation}
\begin{split}
	d s^2 = \eta_{AB}dX^AdX^b &= - \big(d X^0 \big)^2 + d X^a d X_a + \big( d X^{d+1} \big)^2 \\
	&= - d u d r + r^2 d x^a d x_a ,
\end{split}
\end{equation}
where $\eta_{AB} = (-1,1,\ldots,1)$ is the usual Minkowski metric in Cartesian coordinates. The range of the coordinates is
\begin{equation}
\begin{split}
u \in \mrr, \qquad x^a \in \mrr^d , \qquad r \in \mrr\setminus\{0\} . 
\end{split}
\end{equation}
We note
\begin{equation}
\begin{split}
\p_u &= \frac{1}{2} \left[ \p_{X^0} -  \p_{X^{d+1} } \right]  \\
\p_a  &= r x_a \left[  \p_{X^0} - \p_{X^{d+1}} \right] + r \p_{X^a}   \\
\p_r &=  \frac{1}{2} \left( 1 + x^2 \right) \p_{X^0 } + x^a \p_{X^a} + \frac{1}{2} \left( 1 - x^2 \right) \p_{X^{d+1}} , \\
\end{split}
\end{equation}
and
\begin{equation}
\begin{split}
d X^0 &= \frac{1}{2} \left( 1 + x^2 \right) d r + \frac{1}{2} d u + r x_a d x^a  \\
d X^a &= x^a d r + r d x^a  \\
d X^{d+1} &= \frac{1}{2} \left( 1 - x^2 \right) d r - \frac{1}{2} d u - r x_a d x^a .\\
\end{split}
\end{equation}
The volume form is
\begin{equation}
\begin{split}
\ve &= d X^0 \wedge d X^1 \wedge \cdots \wedge d X^d \wedge d X^{d+1}  = \frac{1}{2} r^d d u \wedge d x^1 \wedge \cdots \wedge d x^d \wedge d r .
\end{split}
\end{equation}
The nonvanishing metric components and its inverse are
\begin{equation}
\begin{split}
g_{ur} = - \frac{1}{2} , \qquad g_{ab} = r^2 \delta_{ab} , \qquad g^{ur} = - 2 , \qquad g^{ab} = \frac{1}{r^2} \delta^{ab} . 
\end{split}
\end{equation}
The nonvanishing Christoffel symbols are
\begin{equation}
\begin{split}
\G^u_{ab}[g] = 2 r \delta_{ab} , \qquad \G^a_{rb}[g] = \frac{1}{r} \delta^a_b . 
\end{split}
\end{equation}

\subsubsection{Boundaries of Minkowski Spacetime}

To determine the location of the boundaries of Minkowski spacetime in flat null coordinates, we relate them to Eddington-Finkelstein coordinates $(U,V,\l,\T)$, which are
\begin{equation}
\begin{split}
X^0 = \frac{V+U}{2} , \qquad X^a = \frac{V-U}{2} \sqrt{ 1 - \l^2 } {\hat n}^a ( \T ) , \qquad X^{d+1} = \frac{V-U}{2} \l .
\end{split}
\end{equation}
Note that the radial distance is simply expressed as $R \equiv \sqrt{ X^a X_a + \big(X^{d+1}\big)^2 } = \frac{V-U}{2}$. The coordinate ranges are
\begin{equation}
\begin{split}
V \geq U , \qquad \l \in (-1,1) , \qquad \T \in S^{d-1} . 
\end{split}
\end{equation}
Together, the coordinates $(\l,\T)$ parametrize $S^d$. $\l\in(0,1)$ is the northern hemisphere and $\l\in(-1,0)$ is the southern hemisphere. $\l=0$ is the equator, $\l=+1$ is the north pole and $\l=-1$ is the south pole. Let $\big(\l_\ap,\T_\ap\big)$ be the antipodal point to $(\l,\T)$. Then
\begin{equation}
\begin{split}
{\hat n}\big(\T_\ap\big) = - {\hat n}(\T) , \qquad \l_\ap = - \l . 
\end{split}
\end{equation}
Furthermore, in terms of the flat null coordinates, we have
\begin{equation}
\begin{split}
u &= \frac{ 2 U V } { V ( 1 + \l ) + U ( 1 - \l ) } = \frac{  \big(X^0\big)^2 - R^2  } {   X^0  +   \l R  }   \\
x^a &= \frac{(V-U) \sqrt{1-\l^2} {\hat n}^a(\T)}{ V ( 1 + \l ) + U ( 1 - \l ) }= \frac{ R  \sqrt{1-\l^2} {\hat n}^a(\T)}{ X^0  +   \l R }     \\
r &= \frac{V}{2}(1+\l) + \frac{U}{2}(1-\l) = X^0 + \l R .
\end{split}
\end{equation}
We can now determine the boundaries of Minkowski spacetime in flat null coordinates.

\begin{enumerate}
\item \textbf{Future Null Infinity $\ci^+$:} This is located at $V \to +\infty$ while keeping $(U,\l,\T)$ fixed. In this limit,
\begin{equation}
\begin{split}
u &\to \frac{ 2 U } { 1 + \l } , \qquad x^a \to  \sqrt{ \frac{1-\l}{1+\l} }  {\hat n}^a(\T) , \qquad r \to \frac{V}{2}(1+\l) .
\end{split}
\end{equation}
Thus, we find that $\ci^+$ is located at $r \to \infty$. At this point, $x^a$ are the stereographic coordinates on $S^d$. The north pole ($\lambda=1$) is at $|x^a|=0$, the south pole ($\lambda=-1$) is at $|x^a|=\infty$, and the equator ($\lambda=0$) is on the unit circle $x^a x_a = 1$. $u$ is the null generator of this surface. 

\item \textbf{Past Null Infinity $\ci^-$:} This is located at $U \to -\infty$ while keeping $(V,\l,\T)$ fixed. In this limit,
\begin{equation}
\begin{split}
u &\to \frac{2V}{1 + \l_\ap } , \qquad x^a \to  \sqrt{ \frac{ 1 - \l_\ap }{ 1 + \l_\ap } }  {\hat n}^a\big(\T_\ap\big) , \qquad r \to - \frac{|U|}{2}\left(1+\l_\ap\right).
\end{split}
\end{equation}
Thus, we find that $\ci^-$ is located at $r \to - \infty$. Again, $x^a$ are the stereographic coordinates on $S^d$, but now with an antipodal identification. The north pole ($\lambda=1$) is at $|x^a|=\infty$, the south pole ($\lambda=-1$) is at $|x^a|=0$, and the equator ($\lambda=0$) is on the unit circle $x^a x_a = 1$. $u$ is the null generator of this surface.

\item \textbf{Future Timelike Infinity $i^+$:} This is located at $X^0 \to +\infty$ while keeping $(R,\l,\T)$ fixed. In this limit,
\begin{equation}
\begin{split}
u &\to X^0 - \l R , \qquad x^a \to \frac{R  \sqrt{1-\l^2}}{ X^0 }  {\hat n}^a(\T) , \qquad r \to X^0 + \l R  . 
\end{split}
\end{equation}
Thus, we find that $i^+$ located at $u,r \to \infty$ and $x^a \to 0$ while keeping $u-r$ and $rx^a$ fixed.

\item \textbf{Past Timelike Infinity $i^-$:} This is located at $X^0 \to -\infty$ while keeping $(R,\l,\T)$ fixed. In this limit,
\begin{equation}
\begin{split}
u &\to - \left( \big|X^0\big| - \l_\ap R \right) , \quad x^a \to \frac{R  \sqrt{1-\l_\ap^2} }{ \big|X^0\big| } {\hat n}^a\left(\T_\ap\right) , \quad r \to - \big| X^0\big|  - \l_\ap R  . 
\end{split}
\end{equation}
Thus, we find that $i^-$ located at $u,r \to - \infty$ and $x^a \to 0$ while keeping $u-r$ and $rx^a$ fixed.

\item \textbf{Spatial Infinity $i^0$:} This is located at $R \to +\infty$ while keeping $(X^0,\l,\T)$ fixed. In this limit,
\begin{equation}
\begin{split}
u &\to - \frac{R}{\l} + \frac{X^0}{\l^2} , \qquad x^a \to \frac{1 }{ \l }   \sqrt{1-\l^2} {\hat n}^a(\T)  , \qquad r = \l R  + X^0 . 
\end{split}
\end{equation}
This limit corresponds to $ur\to-\infty$ while keeping $\frac{u}{r}$ and $x^a$ fixed. In particular, on the northern hemisphere, we take $u \to -\infty$ and $r \to + \infty$, whereas on the southern hemisphere, we take $u \to + \infty$ and $r \to - \infty$.

\end{enumerate}

\subsubsection{Isometries}

The isometries of $\mrr^{1,d+1}$ are the $d+2$ translations and the $O(1,d+1)$ rotations. The corresponding Killing vectors in flat null coordinates are
\begin{equation}
\begin{split}
\xi_f &= f(x) \p_u - \frac{1}{2r} \p^a f(x) \p_a + \frac{1}{2d} \p^2 f(x) \p_r   \\
\zeta_Y &= \psi(x) \left( u \p_u - r \p_r \right) + \left[ Y^a(x) - \frac{u}{2r} \p^a \psi(x) \right] \p_a  ,
\end{split}
\end{equation}
where $\psi(x) = \frac{1}{d} \p_a Y^a (x) $, and $f(x)$ and $Y^a(x)$ satisfy the equations
\begin{equation}
\begin{split}
\p_a \p_b f(x) = \frac{1}{d} \delta_{ab} \p^2 f(x) , \qquad \p_a Y_b(x) + \p_b Y_a(x) = \frac{2}{d} \delta_{ab} \p^c Y_c (x) .\\
\end{split}
\end{equation}
This implies that $Y^a(x)$ is the conformal Killing vector of $\mrr^d$ (which manifests the isomorphism $SO(1,d+1) \cong \mathfrak{conf}(\mrr^d)$). The explicit solutions to this are
\begin{equation}
\begin{split}
f(x) = f_0 + f_a x^a + f_{d+1} x^2 , \qquad Y^a(x) = \chi^a + \o^a{}_b x^b + \lambda x^a + \zeta^a x^2 - 2 x^a \zeta \cdot x . 
\end{split}
\end{equation}
Thus, $\psi(x) = \lambda - 2 \zeta \cdot x$. The $SO(d)$ rotations have $\psi(x) = 0$ (generated by $\chi^a$ and $\o^a{}_b$), whereas the boosts in $SO(1,d+1)$ have $\psi(x) \neq 0$ (generated by $\lambda$ and $\zeta^a$). Finally, we remark that the Lie algebra of the Killing vectors (the Poincar\'e algebra) takes the form
\begin{equation}
\begin{split}
\left[ \xi_f , \xi_{f'} \right] &= 0 , \qquad \left[ \zeta_Y , \xi_f \right] = \xi_{Y^a\p_af - f \psi } , \qquad \left[ \zeta_Y , \zeta_{Y'} \right] = \zeta_{[Y,Y']} .
\end{split}
\end{equation}

 \subsection{Momentum Space}\label{sec:mom}

In this subsection, we observe a general parameterization of off-shell momenta and list some useful formulae.

\subsubsection{Off-Shell Momentum}\label{offshellpar}

Following closely the relationship between flat null coordinates and Cartesian coordinates, we parameterize an \emph{off-shell} momentum as
\begin{equation}
\begin{split}
q^0 = \frac{\o}{2} \left( 1 + y^2 \right)  + \frac{\mu}{2\o} , \qquad q^a = \o y^a , \qquad q^{d+1} = \frac{\o}{2} \left( 1 - y^2 \right)  - \frac{\mu}{2\o} .
\end{split}
\end{equation}
Inversely, $\mu =  - q^A q_A$, $y^a = q^a(q^0 + q^{d+1})^{-1}$, and $\o = q^0 + q^{d+1}$. This parameterization is valid everywhere except when $\o=q^0 + q^{d+1} = 0$. The range of these coordinates is $\mu \in \mrr$, $y^a \in \mrr^d$ and $\o \in \mrr\setminus\{0\}$. The momentum space volume form is
\begin{equation}
\begin{split}
\ve &= d q^0 \wedge d q^1 \wedge \cdots \wedge d q^d \wedge d q^{d+1} = \frac{1}{2} \o^{d-1} d \mu \wedge d y^1 \wedge \cdots \wedge d y^d \wedge d \o . 
\end{split}
\end{equation}
This implies that the off-shell integration measure and Dirac delta function are
\begin{equation}
\begin{split}
\int \frac{ d^{d+2} q }{ (2\pi)^{d+2}} &= \frac{1}{2(2\pi)^{d+2}} \int_{-\infty}^\infty d \mu \int_{-\infty}^\infty d \o \int_{\mrr^d} d ^d y  \,|\o|^{d-1}  \\
\delta^{(d+2)} \left( q - q' \right) &= 2|\o|^{1-d} \delta \left( \mu - \mu' \right) \delta \left( \o - \o' \right) \delta^{(d)} \left( y - y' \right) .
\end{split}
\end{equation}

\subsubsection{On-Shell Momentum and Polarization Vectors}\label{onshellpar}

The on-shell condition on a momentum implies
\begin{equation}
\begin{split}
q^A q_A = - m^2, \qquad q^0 > 0 \quad \implies \quad \mu = m^2 \geq 0 , \qquad \o > 0 . 
\end{split}
\end{equation}
The on-shell integration measure and Dirac delta function are
\begin{equation}
\begin{split}
\int \frac{ d^{d+2} q }{ (2\pi)^{d+2}} (2\pi) \delta \big( q^2 + m^2 \big) \t(q^0) &= \int \frac{ d^{d+1} q }{ (2\pi)^{d+1}} \frac{1}{2q^0} = \frac{1}{2(2\pi)^{d+1}}  \int_0^\infty d\o \int_{\mrr^d} d^d y\,\o^{d-1}  \\
 ( 2 q^0  ) \delta^{(d+1)} \left( \vec{q} - \vec{q}\,'\right) &= 2 \o^{1- d}  \delta \left( \o - \o' \right) \delta^{(d)} \left( y - y' \right) .  
\end{split}
\end{equation}

The $d+1$ polarizations of a massive particle are taken to be
\begin{equation}
\begin{split}
\ve_a^A (y) &= \left( y_a , \delta_a^b , - y_a \right) ,  \qquad \ve^A_\o(\o,y) = \frac{1}{2} \left( 1 + y^2 - \frac{m^2}{\o^2} , 2 y^a , 1 - y^2 + \frac{m^2}{\o^2} \right)   .  
\end{split}
\end{equation}
In the massless limit $\ve^A_\o \to \o^{-1} q^A$. Thus, in this limit, this is the pure gauge mode and does not contribute to physical amplitudes. For future reference, we collect the formulae
\begin{equation}
\begin{split}
q ( \o,y) \cdot q ( \o' , y' ) &= - \frac{1}{2} \o \o' \left[ (y - y' )^2 + \frac{m^2}{\o^2} + \frac{m'^2}{\o'^2} \right]   \\
q(\o,y) \cdot \ve_a(y') &= \o ( y - y' )_a   \\ 
q(\o,y) \cdot \ve_{\o'}(\o',y') &= - \frac{\o}{2} \left[ ( y - y' )^2 + \frac{m^2}{\o^2} - \frac{m'^2}{\o'^2}  \right]  \\
\ve_a(y) \cdot \ve_b(y') &= \delta_{ab}  \\
\ve_a(y) \cdot \ve_{\o'}(\o',y') &=  - ( y - y' )_a   \\
\ve_{\o} (\o, y) \cdot \ve_{\o'}(\o',y') &= - \frac{1}{2} \left[   ( y - y' )^2 - \frac{m^2}{\o^2} - \frac{m'^2}{\o'^2}  \right]    \\
X(u,r,x) \cdot q ( \o,y) &= - \frac{\o u}{2} - \frac{\o r}{ 2} \left[ ( x - y )^2 + \frac{m^2}{\o^2} \right]  \\
X(u,r,x) \cdot \ve_a(\o,y) &= r ( x - y )_a  \\
X(u,r,x) \cdot \ve_{\o} (\o,y) &= - \frac{u}{2} - \frac{r}{2} \left[ ( x - y )^2 - \frac{m^2}{\o^2} \right] . \\
\end{split}
\end{equation}
We also require the components of the momentum and polarization vectors in flat null coordinates
\begin{equation}
\begin{split}
q_u ( \o,y) &= - \frac{\o}{2}  \\
q_r (\o,y) &= - \frac{\o}{2} \left[ ( x - y )^2 + \frac{m^2}{\o^2} \right]  \\
q_a (\o,y) &= - r \w ( x - y )_a  \\
(\ve_a)_u(\o,y) &= 0  \\
(\ve_a)_r(\o,y) &= ( x - y )_a  \\
(\ve_a)_b ( \o,y) &= r \delta_{ab}  \\
(\ve_\o)_u ( \o,y) &= - \frac{1}{2}  \\
(\ve_\o)_r ( \o,y)  &= - \frac{1}{2}  \left[  ( x - y )^2 - \frac{m^2}{\o^2} \right]  \\
(\ve_\o)_a ( \o,y)  &= - r ( x - y )_a . 
\end{split}
\end{equation}
For massless vectors, we note from the formulae above that our polarization choice is equivalent to the gauge choice $A_u = 0$. The polarization tensors for the massless field strength then has the following components in flat null coordinates
\begin{equation}
\begin{split}\label{poltensflat}
(\ve_a)_{ur}(\o,y) &= - \frac{i}{2} \o ( x - y )_a  \\
(\ve_a)_{rb} ( \o , y )  &= - \frac{i}{2} \o r ( x - y )^2 \CI_{ab} ( x - y )  \\
(\ve_a)_{ub}(\o,y) &= - \frac{i}{2} \o r \delta_{ab}  \\
(\ve_c)_{ab} ( \o,y) &= - i \o r^2 \left[ ( x - y )_a \delta_{bc} - ( x - y )_b \delta_{ac} \right] , 
\end{split}
\end{equation}
where $\CI_{ab}(x-y)$ is the conformally invariant tensor
\begin{equation}
\begin{split}
\CI_{ab}(x-y) = \delta_{ab} - \frac{2 ( x - y )_a ( x - y )_b }{ (  x - y )^2 }  .
\end{split}
\end{equation}
 
 \section{Action of the Hard Charge}\label{app:scalar}
 
In this section, we show that the hard charge generates large gauge transformations on the matter states by an explicit computation for a minimally coupled scalar field $\Phi$ with $U(1)$ charge $Q \in \mzz$. The corresponding conserved current is 
\begin{equation}
\begin{split}
J_{\mu} = i Q \left( \Phi^* D_\mu \Phi - ( D_\mu  \Phi )^* \Phi \right) , \qquad D_\mu \Phi = \p_\mu \Phi - i Q A_\mu \Phi . 
\end{split}
\end{equation}
The product of fields is defined via normal ordering.

The mode expansion for the outgoing/incoming scalar field is
\begin{equation}
\begin{split} 
\Phi^\pmm(X) =   \int \frac{d^{d+1} q}{ ( 2\pi )^{d+1} } \frac{1}{2q^0} \left[  \CO_{\Phi}^\pmm(\vec{q}\,) e^{ i q \cdot X }  + \CO_{ {\overline \Phi}}^{\pmm\dagger} (\vec{q}\,) e^{ - i q \cdot X }  \right] ,
\end{split}
\end{equation}
where
\begin{equation}
\begin{split}
\left[ \CO_{\Phi}^\pmm (\vec{q}\,) , \CO_{\Phi}^{\pmm\dagger} (\vec{q}\,') \right] = \left[ \CO_{ {\overline \Phi}}^\pmm (\vec{q}\,) , \CO_{ {\overline \Phi}}^{\pmm\dagger} (\vec{q}\,') \right]  = ( 2 q^0  ) (2\pi)^{d+1}  \delta^{(d+1)} \left( \vec{q} - \vec{q}\,' \right) .
\end{split}
\end{equation}
Moving to flat null coordinates and using \eqref{mompar}, we find
\begin{equation}
\begin{split} 
\Phi^\pmm(u,r,x) &=  \frac{1}{2(2\pi)^{d+1}}  \int_{\mrr^d} d^d y  \int_0^\infty d\o \,\o^{d-1}  \left[  \CO_{\Phi}^\pmm(\o,x+y) e^{- \frac{i}{2} \o u - \frac{i}{2} \o r y^2  }  \right. \\
& \left. \qquad \qquad \qquad \qquad \qquad \qquad \qquad \qquad \qquad + \CO_{ {\overline \Phi}}^{\pmm\dagger} (\o,x+y) e^{ \frac{i}{2} \o u + \frac{i}{2} \o r y^2  }  \right] ,
\end{split}
\end{equation}
where
\begin{equation}
\begin{split}\label{scalarcommrel}
\left[ \CO_{\Phi}^\pmm (\o,x) , \CO_{\Phi}^{\pmm\dagger} (\o',x') \right] &= \left[ \CO_{ {\overline \Phi}}^\pmm (\o,x) , \CO_{ {\overline \Phi}}^{\pmm\dagger} (\o',x') \right] \\
&= 2\o^{1-d} (2\pi)^{d+1} \delta(\o-\o') \delta^{(d)} \left( x - x'  \right) .
\end{split}
\end{equation}

The leading order term in the large $r$ expansion of the scalar field is (this is determined following precisely the method of \S\ref{sec:radasymptotics})
\begin{equation}
\begin{split}
	\Phi^\pmm ~~\to~~  \frac{1}{2(2\pi)^{\frac{d}{2}+1}} \int_0^\infty d\o \,\o^{\frac{d}{2}-1}  \left[ \frac{e^{- \frac{i}{2} \o u }}{(ir)^{\frac{d}{2}}}  \CO_{\Phi}^\pmm(\o,x)   + \frac{e^{ \frac{i}{2} \o u   }}{(-ir)^{\frac{d}{2}}} \CO_{ {\overline \Phi}}^{\pmm\dagger} (\o,x)   \right] + \cdots  .
\end{split}
\end{equation}
We then determine that 
\begin{equation}
\begin{split}\label{scalarcurr}
\int du\, J_u^{(\pm,d)} &=  \frac{ \pi Q  }{(2\pi)^{d+2} } \int_0^\infty d\o \,\o^{d-1} \left[  \CO_{\Phi}^{\pmm\dagger}(\o,x)  \CO_{\Phi}^\pmm(\o,x)  - \CO_{ {\overline \Phi}}^{\pmm\dagger} (\o,x)\CO_{ {\overline \Phi}}^{\pmm} (\o,x) \right]  . 
\end{split}
\end{equation}
An outgoing or incoming scalar state of charge $Q$ is
\begin{equation}
\begin{split}
\bra{\Phi,\o , x} = \bra{0} \CO_{\Phi}^{\+}(\o,x)   , \qquad \ket{\Phi,\o , x } =  \CO_{\Phi}^{\-\dagger}(\o,x)  \ket{0} . 
\end{split}
\end{equation}
Using these definitions, we find 
\begin{equation}
\begin{split}
\bra{\Phi,\o , x' } \int du \,J_u^{(+ ,d)} (u,x) &= Q  \delta^{(d)}(x - x' ) \bra{\Phi,\o , x' }   \\
  \int du\, J_u^{(-,d)}(u,x) \ket{\Phi,\o , x' }  &=  Q  \delta^{(d)}(x - x' ) \ket{\Phi,\o , x' } . 
\end{split}
\end{equation}
The action of the hard charge \eqref{hardcharge} is then
\begin{equation}
\begin{split}
\bra{\Phi,\o , x } Q^{+H}_\ve = - Q \ve(x)\bra{\Phi,\o , x } , \qquad Q^{-H}_\ve \ket{\Phi,\o , x }   = - Q \ve(x)\ket{\Phi,\o , x }   .
\end{split}
\end{equation}

\section{Soft Photon Operator}\label{softinsertion}

We describe how to obtain the soft photon insertion operator in an $S$-matrix using the path integral approach. First, the $n$-point amplitude is
\begin{equation}
\begin{split}
\CA_n ( p_1 , \cdots , p_n  )  = \braket{\text{out}}{\text{in}} . 
\end{split}
\end{equation}
This is computed in the path integral formalism by computing off-shell momentum space correlators and taking an appropriate on-shell limit. For instance, for an outgoing scalar particle with momentum $\vec{p}$, we insert
\begin{equation}
\begin{split}
i Z_\Phi \lim_{p^0 \to \sqrt{ | \vec{p}\,| + m^2 } - i \e } \left( p^2 - i \e \right) \int d^{d+2} X \, e^{- i p \cdot X } \Phi(X)  . 
\end{split}
\end{equation}
Here, $\Phi(X)$ is an operator that creates (among other things) a scalar one-particle state, and $Z_\Phi$ is the corresponding wave-function renormalization factor. 

The amplitude involving an extra outgoing photon, i.e. $\CA_{n+1}^\text{out}$ from \eqref{softthmmom}, similarly corresponds to the insertion of the operator
\begin{equation}
\begin{split}
\frac{i Z}{e} \lim_{p_\g^0 \to |\vec{p}_\g| - i \e } \left( p_\g^2 - i \e \right)  \ve_a^A ( \vec{p}_\g)^* \int d^{d+2} X \, e^{ - i p_\g \cdot X }   A_A (X)  . 
\end{split}
\end{equation}
To understand what this operator is in terms of the $in$ and $out$ creation/annihilation operators, we expand the gauge field $A_A(x)$ in terms of time-dependent mode coefficients:
\begin{equation}
\begin{split}
A_A (X) = e \int \frac{d^{d+1} q}{ ( 2\pi )^{d+1} } \frac{1}{2q^0} \left[ \ve^a_{A} ({\vec q}\,) \CO_a(X^0,\vec{q}\,) e^{ i q \cdot X }  +  \ve^a_{A} ({\vec q}\,)^* \CO^\dagger_a(X^0,\vec{q}\,) e^{ - i q \cdot X }  \right] .
\end{split}
\end{equation}
Using this, we find
\begin{equation}\label{insertionop}
\begin{split}
\frac{iZ}{e} \lim_{p_\g^0 \to |\vec{p}_\g| - i \e } \left( p_\g^2 - i \e \right)  \ve_a^A ( \vec{p}_\g)^* \int d^{d+2} X  \, e^{ - i p_\g \cdot X }   A_A (X)  = \CO^\+_a(\vec{p}_\g) - \CO^{\-}_a(\vec{p}_\g) .
\end{split}
\end{equation}
where $\lim_{X^0 \to \pm \infty} \CO_a ( X^0 , \vec{p}_\g ) = Z^{-1} \CO_a^\pmm(\vec{p}_\g)$.\footnote{To show this, we additionally assume that $\lim\limits_{X^0 \to \pm \infty} \p_0 \CO_a ( X^0 , \vec{p}_\g ) = 0$ (see Schwartz \cite{Schwartz:2013pla} for a discussion on this).} For photons with non-zero energy, the operator $\CO^{\-}_a(\vec{p}_\g)$ annihilates the $in$ vacuum (assuming there are no other incoming photons with the same momentum, i.e. assuming no forward scattering) so that $S$-matrices with energetic outgoing photons correspond to insertions of $\CO^\+_a(\vec{p}_\g)$. On the other hand, owing to \eqref{modesconstraint}, $\CO_a^{(-,0)}(x)$ no longer annihilates the vacuum; rather, it creates an incoming soft photon (to be more precise, it produces a new vacuum state). Therefore, it follows that the operator generating the soft limit must include both terms from \eqref{insertionop} and is given by
\begin{equation}
\begin{split}
	\lim_{\o \to 0}  \left[ \o\CO^\+_a(\o,x) - \o\CO^{\-}_a(\o,x) \right] = \CO_a^{(+,0)}(x) -  \CO_a^{(-,0)}(x) . 
\end{split}
\end{equation}

\bibliography{HM-bib}{}

\providecommand{\href}[2]{#2}\begingroup\raggedright\begin{thebibliography}{10}

\bibitem{Bloch:1937pw}
F.~Bloch and A.~Nordsieck, ``{Note on the Radiation Field of the electron},''
\href{http://dx.doi.org/10.1103/PhysRev.52.54}{{\em Phys. Rev.} {\bfseries 52}
  (1937) 54--59}.

\bibitem{Nordsieck:1937zz}
A.~Nordsieck, ``{The Low Frequency Radiation of a Scattered Electron},''
\href{http://dx.doi.org/10.1103/PhysRev.52.59}{{\em Phys. Rev.} {\bfseries 52}
  (1937) 59--62}.

\bibitem{Low:1954kd}
F.~E. Low, ``{Scattering of light of very low frequency by systems of spin
  1/2},''
\href{http://dx.doi.org/10.1103/PhysRev.96.1428}{{\em Phys. Rev.} {\bfseries
  96} (1954) 1428--1432}.

\bibitem{GellMann:1954kc}
M.~Gell-Mann and M.~L. Goldberger, ``{Scattering of low-energy photons by
  particles of spin 1/2},''
\href{http://dx.doi.org/10.1103/PhysRev.96.1433}{{\em Phys. Rev.} {\bfseries
  96} (1954) 1433--1438}.

\bibitem{Low:1958sn}
F.~E. Low, ``{Bremsstrahlung of very low-energy quanta in elementary particle
  collisions},''
\href{http://dx.doi.org/10.1103/PhysRev.110.974}{{\em Phys. Rev.} {\bfseries
  110} (1958) 974--977}.

\bibitem{Yennie:1961ad}
D.~R. Yennie, S.~C. Frautschi, and H.~Suura, ``{The infrared divergence
  phenomena and high-energy processes},''
\href{http://dx.doi.org/10.1016/0003-4916(61)90151-8}{{\em Annals Phys.}
  {\bfseries 13} (1961) 379--452}.

\bibitem{Weinberg:1965nx}
S.~Weinberg, ``{Infrared photons and gravitons},''
\href{http://dx.doi.org/10.1103/PhysRev.140.B516}{{\em Phys.Rev.} {\bfseries
  140} (1965) B516--B524}.

\bibitem{Weinberg:1995mt}
S.~Weinberg, {\em {The Quantum theory of fields. Vol. 1: Foundations}}.
\newblock Cambridge University Press,
2005.
\newblock

\bibitem{Strominger:2017zoo}
A.~Strominger, ``{Lectures on the Infrared Structure of Gravity and Gauge
  Theory},''
\href{http://arxiv.org/abs/1703.05448}{{\ttfamily arXiv:1703.05448 [hep-th]}}.

\bibitem{He:2014cra}
T.~He, P.~Mitra, A.~P. Porfyriadis, and A.~Strominger, ``{New Symmetries of
  Massless QED},'' \href{http://dx.doi.org/10.1007/JHEP10(2014)112}{{\em JHEP}
  {\bfseries 10} (2014) 112},
\href{http://arxiv.org/abs/1407.3789}{{\ttfamily arXiv:1407.3789 [hep-th]}}.

\bibitem{Mohd:2014oja}
A.~Mohd, ``{A note on asymptotic symmetries and soft-photon theorem},''
  \href{http://dx.doi.org/10.1007/JHEP02(2015)060}{{\em JHEP} {\bfseries 02}
  (2015) 060},
\href{http://arxiv.org/abs/1412.5365}{{\ttfamily arXiv:1412.5365 [hep-th]}}.

\bibitem{He:2015zea}
T.~He, P.~Mitra, and A.~Strominger, ``{2D Kac-Moody Symmetry of 4D Yang-Mills
  Theory},'' \href{http://dx.doi.org/10.1007/JHEP10(2016)137}{{\em JHEP}
  {\bfseries 10} (2016) 137},
\href{http://arxiv.org/abs/1503.02663}{{\ttfamily arXiv:1503.02663 [hep-th]}}.

\bibitem{Campiglia:2015qka}
M.~Campiglia and A.~Laddha, ``{Asymptotic symmetries of QED and Weinberg's soft
  photon theorem},'' \href{http://dx.doi.org/10.1007/JHEP07(2015)115}{{\em
  JHEP} {\bfseries 07} (2015) 115},
\href{http://arxiv.org/abs/1505.05346}{{\ttfamily arXiv:1505.05346 [hep-th]}}.

\bibitem{Kapec:2015ena}
D.~Kapec, M.~Pate, and A.~Strominger, ``{New Symmetries of QED},''
  \href{http://dx.doi.org/10.4310/ATMP.2017.v21.n7.a7}{{\em Adv. Theor. Math.
  Phys.} {\bfseries 21} (2017) 1769--1785},
\href{http://arxiv.org/abs/1506.02906}{{\ttfamily arXiv:1506.02906 [hep-th]}}.

\bibitem{Strominger:2015bla}
A.~Strominger, ``{Magnetic Corrections to the Soft Photon Theorem},''
  \href{http://dx.doi.org/10.1103/PhysRevLett.116.031602}{{\em Phys. Rev.
  Lett.} {\bfseries 116} no.~3, (2016) 031602},
\href{http://arxiv.org/abs/1509.00543}{{\ttfamily arXiv:1509.00543 [hep-th]}}.

\bibitem{Gabai:2016kuf}
B.~Gabai and A.~Sever, ``{Large gauge symmetries and asymptotic states in
  QED},'' \href{http://dx.doi.org/10.1007/JHEP12(2016)095}{{\em JHEP}
  {\bfseries 12} (2016) 095},
\href{http://arxiv.org/abs/1607.08599}{{\ttfamily arXiv:1607.08599 [hep-th]}}.

\bibitem{He:2014laa}
T.~He, V.~Lysov, P.~Mitra, and A.~Strominger, ``{BMS supertranslations and
  Weinberg's soft graviton theorem},''
  \href{http://dx.doi.org/10.1007/JHEP05(2015)151}{{\em JHEP} {\bfseries 05}
  (2015) 151},
\href{http://arxiv.org/abs/1401.7026}{{\ttfamily arXiv:1401.7026 [hep-th]}}.

\bibitem{Campiglia:2015yka}
M.~Campiglia and A.~Laddha, ``{New symmetries for the Gravitational
  S-matrix},'' \href{http://dx.doi.org/10.1007/JHEP04(2015)076}{{\em JHEP}
  {\bfseries 04} (2015) 076},
\href{http://arxiv.org/abs/1502.02318}{{\ttfamily arXiv:1502.02318 [hep-th]}}.

\bibitem{Campiglia:2015kxa}
M.~Campiglia and A.~Laddha, ``{Asymptotic symmetries of gravity and soft
  theorems for massive particles},''
  \href{http://dx.doi.org/10.1007/JHEP12(2015)094}{{\em JHEP} {\bfseries 12}
  (2015) 094},
\href{http://arxiv.org/abs/1509.01406}{{\ttfamily arXiv:1509.01406 [hep-th]}}.

\bibitem{Dumitrescu:2015fej}
T.~T. Dumitrescu, T.~He, P.~Mitra, and A.~Strominger, ``{Infinite-Dimensional
  Fermionic Symmetry in Supersymmetric Gauge Theories},''
\href{http://arxiv.org/abs/1511.07429}{{\ttfamily arXiv:1511.07429 [hep-th]}}.

\bibitem{Cachazo:2014fwa}
F.~Cachazo and A.~Strominger, ``{Evidence for a New Soft Graviton Theorem},''
\href{http://arxiv.org/abs/1404.4091}{{\ttfamily arXiv:1404.4091 [hep-th]}}.

\bibitem{Schwab:2014fia}
B.~U.~W. Schwab, ``{Subleading Soft Factor for String Disk Amplitudes},''
\href{http://arxiv.org/abs/1406.4172}{{\ttfamily arXiv:1406.4172 [hep-th]}}.

\bibitem{Kalousios:2014uva}
C.~Kalousios and F.~Rojas, ``{Next to subleading soft-graviton theorem in
  arbitrary dimensions},''
  \href{http://dx.doi.org/10.1007/JHEP01(2015)107}{{\em JHEP} {\bfseries 01}
  (2015) 107},
\href{http://arxiv.org/abs/1407.5982}{{\ttfamily arXiv:1407.5982 [hep-th]}}.

\bibitem{Luo:2014wea}
H.~Luo, P.~Mastrolia, and W.~J. Torres~Bobadilla, ``{Subleading soft behavior
  of QCD amplitudes},''
  \href{http://dx.doi.org/10.1103/PhysRevD.91.065018}{{\em Phys. Rev.}
  {\bfseries D91} no.~6, (2015) 065018},
\href{http://arxiv.org/abs/1411.1669}{{\ttfamily arXiv:1411.1669 [hep-th]}}.

\bibitem{DiVecchia:2016amo}
P.~Di~Vecchia, R.~Marotta, and M.~Mojaza, ``{Subsubleading soft theorems of
  gravitons and dilatons in the bosonic string},''
  \href{http://dx.doi.org/10.1007/JHEP06(2016)054}{{\em JHEP} {\bfseries 06}
  (2016) 054},
\href{http://arxiv.org/abs/1604.03355}{{\ttfamily arXiv:1604.03355 [hep-th]}}.

\bibitem{Sen:2017nim}
A.~Sen, ``{Subleading Soft Graviton Theorem for Loop Amplitudes},''
\href{http://arxiv.org/abs/1703.00024}{{\ttfamily arXiv:1703.00024 [hep-th]}}.

\bibitem{Laddha:2017vfh}
A.~Laddha and P.~Mitra, ``{Asymptotic Symmetries and Subleading Soft Photon
  Theorem in Effective Field Theories},''
  \href{http://dx.doi.org/10.1007/JHEP05(2018)132}{{\em JHEP} {\bfseries 05}
  (2018) 132},
\href{http://arxiv.org/abs/1709.03850}{{\ttfamily arXiv:1709.03850 [hep-th]}}.

\bibitem{Hirai:2018ijc}
H.~Hirai and S.~Sugishita, ``{Conservation Laws from Asymptotic Symmetry and
  Subleading Charges in QED},''
  \href{http://dx.doi.org/10.1007/JHEP07(2018)122}{{\em JHEP} {\bfseries 07}
  (2018) 122},
\href{http://arxiv.org/abs/1805.05651}{{\ttfamily arXiv:1805.05651 [hep-th]}}.

\bibitem{Barnich:2011ct}
G.~Barnich and C.~Troessaert, ``{Supertranslations call for superrotations},''
  {\em PoS} (2010) 010, \href{http://arxiv.org/abs/1102.4632}{{\ttfamily
  arXiv:1102.4632 [gr-qc]}}.
[Ann. U. Craiova Phys.21,S11(2011)].

\bibitem{Kapec:2014opa}
D.~Kapec, V.~Lysov, S.~Pasterski, and A.~Strominger, ``{Semiclassical Virasoro
  symmetry of the quantum gravity $ \mathcal{S}$-matrix},''
  \href{http://dx.doi.org/10.1007/JHEP08(2014)058}{{\em JHEP} {\bfseries 08}
  (2014) 058},
\href{http://arxiv.org/abs/1406.3312}{{\ttfamily arXiv:1406.3312 [hep-th]}}.

\bibitem{Campiglia:2014yka}
M.~Campiglia and A.~Laddha, ``{Asymptotic symmetries and subleading soft
  graviton theorem},'' \href{http://dx.doi.org/10.1103/PhysRevD.90.124028}{{\em
  Phys. Rev.} {\bfseries D90} no.~12, (2014) 124028},
\href{http://arxiv.org/abs/1408.2228}{{\ttfamily arXiv:1408.2228 [hep-th]}}.

\bibitem{Lysov:2014csa}
V.~Lysov, S.~Pasterski, and A.~Strominger, ``{Low's Subleading Soft Theorem as
  a Symmetry of QED},''
  \href{http://dx.doi.org/10.1103/PhysRevLett.113.111601}{{\em Phys. Rev.
  Lett.} {\bfseries 113} no.~11, (2014) 111601},
\href{http://arxiv.org/abs/1407.3814}{{\ttfamily arXiv:1407.3814 [hep-th]}}.

\bibitem{Campiglia:2016jdj}
M.~Campiglia and A.~Laddha, ``{Sub-subleading soft gravitons: New symmetries of
  quantum gravity?},''
  \href{http://dx.doi.org/10.1016/j.physletb.2016.11.046}{{\em Phys. Lett.}
  {\bfseries B764} (2017) 218--221},
\href{http://arxiv.org/abs/1605.09094}{{\ttfamily arXiv:1605.09094 [gr-qc]}}.

\bibitem{Campiglia:2016hvg}
M.~Campiglia and A.~Laddha, ``{Subleading soft photons and large gauge
  transformations},'' \href{http://dx.doi.org/10.1007/JHEP11(2016)012}{{\em
  JHEP} {\bfseries 11} (2016) 012},
\href{http://arxiv.org/abs/1605.09677}{{\ttfamily arXiv:1605.09677 [hep-th]}}.

\bibitem{Conde:2016csj}
E.~Conde and P.~Mao, ``{Remarks on asymptotic symmetries and the subleading
  soft photon theorem},''
  \href{http://dx.doi.org/10.1103/PhysRevD.95.021701}{{\em Phys. Rev.}
  {\bfseries D95} no.~2, (2017) 021701},
\href{http://arxiv.org/abs/1605.09731}{{\ttfamily arXiv:1605.09731 [hep-th]}}.

\bibitem{Campiglia:2016efb}
M.~Campiglia and A.~Laddha, ``{Sub-subleading soft gravitons and large
  diffeomorphisms},'' \href{http://dx.doi.org/10.1007/JHEP01(2017)036}{{\em
  JHEP} {\bfseries 01} (2017) 036},
\href{http://arxiv.org/abs/1608.00685}{{\ttfamily arXiv:1608.00685 [gr-qc]}}.

\bibitem{Conde:2016rom}
E.~Conde and P.~Mao, ``{BMS Supertranslations and Not So Soft Gravitons},''
\href{http://arxiv.org/abs/1612.08294}{{\ttfamily arXiv:1612.08294 [hep-th]}}.

\bibitem{Laddha:2017ygw}
A.~Laddha and A.~Sen, ``{Sub-subleading Soft Graviton Theorem in Generic
  Theories of Quantum Gravity},''
\href{http://arxiv.org/abs/1706.00759}{{\ttfamily arXiv:1706.00759 [hep-th]}}.

\bibitem{Schwab:2014xua}
B.~U.~W. Schwab and A.~Volovich, ``{Subleading Soft Theorem in Arbitrary
  Dimensions from Scattering Equations},''
  \href{http://dx.doi.org/10.1103/PhysRevLett.113.101601}{{\em Phys. Rev.
  Lett.} {\bfseries 113} no.~10, (2014) 101601},
\href{http://arxiv.org/abs/1404.7749}{{\ttfamily arXiv:1404.7749 [hep-th]}}.

\bibitem{Kapec:2014zla}
D.~Kapec, V.~Lysov, and A.~Strominger, ``{Asymptotic Symmetries of Massless QED
  in Even Dimensions},''
\href{http://arxiv.org/abs/1412.2763}{{\ttfamily arXiv:1412.2763 [hep-th]}}.

\bibitem{Kapec:2015vwa}
D.~Kapec, V.~Lysov, S.~Pasterski, and A.~Strominger, ``{Higher-Dimensional
  Supertranslations and Weinberg's Soft Graviton Theorem},''
\href{http://arxiv.org/abs/1502.07644}{{\ttfamily arXiv:1502.07644 [gr-qc]}}.

\bibitem{Thorne:1992sdb}
K.~S. Thorne, ``{Gravitational-wave bursts with memory: The Christodoulou
  effect},''
\href{http://dx.doi.org/10.1103/PhysRevD.45.520}{{\em Phys. Rev.} {\bfseries
  D45} no.~2, (1992) 520--524}.

\bibitem{Strominger:2014pwa}
A.~Strominger and A.~Zhiboedov, ``{Gravitational Memory, BMS Supertranslations
  and Soft Theorems},'' \href{http://dx.doi.org/10.1007/JHEP01(2016)086}{{\em
  JHEP} {\bfseries 01} (2016) 086},
\href{http://arxiv.org/abs/1411.5745}{{\ttfamily arXiv:1411.5745 [hep-th]}}.

\bibitem{Pasterski:2015zua}
S.~Pasterski, ``{Asymptotic Symmetries and Electromagnetic Memory},''
\href{http://arxiv.org/abs/1505.00716}{{\ttfamily arXiv:1505.00716 [hep-th]}}.

\bibitem{Bieri:2015yia}
L.~Bieri, D.~Garfinkle, and S.-T. Yau, ``{Gravitational Waves and Their Memory
  in General Relativity},''
\href{http://arxiv.org/abs/1505.05213}{{\ttfamily arXiv:1505.05213 [gr-qc]}}.

\bibitem{Susskind:2015hpa}
L.~Susskind, ``{Electromagnetic Memory},''
\href{http://arxiv.org/abs/1507.02584}{{\ttfamily arXiv:1507.02584 [hep-th]}}.

\bibitem{Hotta:2016qtv}
M.~Hotta, J.~Trevison, and K.~Yamaguchi, ``{Gravitational Memory Charges of
  Supertranslation and Superrotation on Rindler Horizons},''
  \href{http://dx.doi.org/10.1103/PhysRevD.94.083001}{{\em Phys. Rev.}
  {\bfseries D94} no.~8, (2016) 083001},
\href{http://arxiv.org/abs/1606.02443}{{\ttfamily arXiv:1606.02443 [gr-qc]}}.

\bibitem{Hollands:2016oma}
S.~Hollands, A.~Ishibashi, and R.~M. Wald, ``{BMS Supertranslations and Memory
  in Four and Higher Dimensions},''
\href{http://arxiv.org/abs/1612.03290}{{\ttfamily arXiv:1612.03290 [gr-qc]}}.

\bibitem{Pate:2017vwa}
M.~Pate, A.-M. Raclariu, and A.~Strominger, ``{Color Memory: A Yang-Mills
  Analog of Gravitational Wave Memory},''
  \href{http://dx.doi.org/10.1103/PhysRevLett.119.261602}{{\em Phys. Rev.
  Lett.} {\bfseries 119} no.~26, (2017) 261602},
\href{http://arxiv.org/abs/1707.08016}{{\ttfamily arXiv:1707.08016 [hep-th]}}.

\bibitem{Pate:2017fgt}
M.~Pate, A.-M. Raclariu, and A.~Strominger, ``{Gravitational Memory in Higher
  Dimensions},'' \href{http://dx.doi.org/10.1007/JHEP06(2018)138}{{\em JHEP}
  {\bfseries 06} (2018) 138},
\href{http://arxiv.org/abs/1712.01204}{{\ttfamily arXiv:1712.01204 [hep-th]}}.

\bibitem{Ball:2018prg}
A.~Ball, M.~Pate, A.-M. Raclariu, A.~Strominger, and R.~Venugopalan,
  ``{Measuring Color Memory in a Color Glass Condensate at Electron-Ion
  Colliders},''
\href{http://arxiv.org/abs/1805.12224}{{\ttfamily arXiv:1805.12224 [hep-ph]}}.

\bibitem{Balazs1955}
N.~L~Balazs, ``Wave propagation in even and odd dimensional spaces,''
  \href{http://dx.doi.org/10.1088/0370-1298/68/6/307}{{\em Proceedings of the
  Physical Society. Section A} {\bfseries 68} (1955) \,521}.

\bibitem{Soodak1993}
H.~Soodak and M.~S. Tiersten, ``Wakes and waves in n dimensions,''
  \href{http://dx.doi.org/10.1119/1.17230}{{\em American Journal of Physics}
  {\bfseries 61} no.~5, (1993) 395--401},
  \href{http://arxiv.org/abs/https://doi.org/10.1119/1.17230}{{\ttfamily
  https://doi.org/10.1119/1.17230}}. \url{https://doi.org/10.1119/1.17230}.

\bibitem{Kapec:2017gsg}
D.~Kapec and P.~Mitra, ``{A $d$-Dimensional Stress Tensor for Mink$_{d+2}$
  Gravity},'' \href{http://dx.doi.org/10.1007/JHEP05(2018)186}{{\em JHEP}
  {\bfseries 05} (2018) 186},
\href{http://arxiv.org/abs/1711.04371}{{\ttfamily arXiv:1711.04371 [hep-th]}}.

\bibitem{Campiglia:2017mua}
M.~Campiglia and R.~Eyheralde, ``{Asymptotic $U(1)$ charges at spatial
  infinity},''
\href{http://arxiv.org/abs/1703.07884}{{\ttfamily arXiv:1703.07884 [hep-th]}}.

\bibitem{Prabhu:2018gzs}
K.~Prabhu, ``{Conservation of asymptotic charges from past to future null
  infinity: Maxwell fields},''
  \href{http://dx.doi.org/10.1007/JHEP10(2018)113}{{\em JHEP} {\bfseries 10}
  (2018) 113},
\href{http://arxiv.org/abs/1808.07863}{{\ttfamily arXiv:1808.07863 [gr-qc]}}.

\bibitem{Yoshida:2017fao}
D.~Yoshida and J.~Soda, ``{Electromagnetic Memory Effect Induced by Axion Dark
  Matter},''
\href{http://arxiv.org/abs/1704.04169}{{\ttfamily arXiv:1704.04169 [gr-qc]}}.

\bibitem{Wald:1999wa}
R.~M. Wald and A.~Zoupas, ``{A General definition of `conserved quantities' in
  general relativity and other theories of gravity},''
  \href{http://dx.doi.org/10.1103/PhysRevD.61.084027}{{\em Phys. Rev.}
  {\bfseries D61} (2000) 084027},
\href{http://arxiv.org/abs/gr-qc/9911095}{{\ttfamily arXiv:gr-qc/9911095
  [gr-qc]}}.

\bibitem{Esmaeili:2019hom}
E.~Esmaeili, ``{Asymptotic Symmetries of Maxwell Theory in Arbitrary Dimensions
  at Spatial Infinity},''
\href{http://arxiv.org/abs/1902.02769}{{\ttfamily arXiv:1902.02769 [hep-th]}}.

\bibitem{Schwartz:2013pla}
M.~D. Schwartz, {\em {Quantum Field Theory and the Standard Model}}.
\newblock Cambridge University Press,
2014.
\newblock

\end{thebibliography}\endgroup
\bibliographystyle{utphys}

\end{document}